%% file: main.tex
\documentclass[final]{ieeetmlcn}
\input{content/def}

\def\BibTeX{{\rm B\kern-.05em{\sc i\kern-.025em b}\kern-.08em
    T\kern-.1667em\lower.7ex\hbox{E}\kern-.125emX}}
\AtBeginDocument{\definecolor{tmlcncolor}{cmyk}{0.93,0.59,0.15,0.02}\definecolor{NavyBlue}{RGB}{0,86,125}}

\def\OJlogo{\vspace{-4pt}\includegraphics[height=18pt]{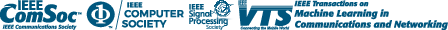}}
\def\seclogo{\vspace{10pt}\includegraphics[height=18pt]{new_logo}}

\def\authorrefmark#1{\ensuremath{^{\textbf{#1}}}}

\begin{document}
\receiveddate{XX Month, XXXX}
\reviseddate{XX Month, XXXX}
\accepteddate{XX Month, XXXX}
\publisheddate{XX Month, XXXX}

\markboth{Evaluation Metrics and Methods for Generative Models in the Wireless PHY Layer}{Baur \textit{et al.}}

\title{Evaluation Metrics and Methods \\for Generative Models in the \\Wireless PHY Layer}

\author{Michael Baur\authorrefmark{1}, Graduate Student Member, IEEE, Nurettin Turan\authorrefmark{1}, Graduate Student Member, IEEE, Simon Wallner\authorrefmark{1}, and Wolfgang Utschick\authorrefmark{1}, Fellow, IEEE}
\affil{TUM School of Computation, Information and Technology, Technical University of Munich, Germany}
\corresp{Corresponding author: Michael Baur (email: mi.baur@tum.de).}
\authornote{M. Baur and N. Turan contributed equally to this work. This work is funded by the Bavarian Ministry of Economic Affairs, Regional Development, and Energy as part of the project 6G Future Lab Bavaria. The authors acknowledge the financial support from the Federal Ministry of Education and Research of Germany, project ID: 16KISK002.}

\begin{abstract}
    \input{content/abs}
\end{abstract}



\maketitle

\input{content/introduction}
\input{content/preliminaries}
\input{content/evaluation_metrics}
\input{content/results}
\input{content/conclusion}
\appendix
\input{content/app}

\balance
\bibliographystyle{IEEEtran}
\bibliography{main}

\vfill\pagebreak

\end{document}

%% file: content/def.tex
\usepackage{tikz}
\usepackage{pgfplots}
\usepackage{bm}
\usepackage{amsmath,amssymb,amsfonts,amsthm}
\usepackage{siunitx}
\usepackage{color}
\usepackage{array,multirow}
\usepackage{adjustbox}
\usepackage{balance}
\usepackage{cite}
\usepackage[acronym,shortcuts]{glossaries}
\usepackage{xspace}
\usepackage{tumcolor}
\usepackage{pgfplotstable}	
\usetikzlibrary{calc, shapes.geometric, arrows, positioning, spy, decorations.fractals, shapes.misc}
\usepackage{algorithm}
\usepackage{algpseudocode}
\usepackage{graphicx,color}
\usepackage{textcomp}
\usepackage{xcolor}
\usepackage{hyperref}
\hypersetup{hidelinks=true}
\usepackage{cleveref}
\usepackage{lipsum}
\usepackage{subfigure}
\usepackage{multirow}
\usepackage{tabularx}

\newcommand\scalemath[2]{\scalebox{#1}{\mbox{\ensuremath{\displaystyle #2}}}}
\pgfplotsset{
	discard if not/.style 2 args={
		x filter/.code={
			\edef\tempa{\thisrow{#1}}
			\edef\tempb{#2}
			\ifx\tempa\tempb
			\else
			\fi
		}
	}
}
\pgfplotsset{compat=1.18}
\definecolor{mylila}{RGB}{153,50,204} 
\definecolor{mygreen}{RGB}{176,191,26} 
\definecolor{ourdarkblue}{RGB}{30, 100, 200}
\definecolor{ourdarkgreen}{RGB}{0, 100, 0}
\definecolor{ouryellow}{RGB}{220, 210, 50}

\def\pltw{245pt}
\def\plth{140pt}
\def\hstw{260pt}
\def\hsth{140pt}
\def\tblw{230pt}

\def\markSize{1.8pt}
\def\lineWidth{1.5pt}

\def\histWidth{0.04cm}

\tikzset{RPE/.style={mark options=solid, color=TUMBeamerBlue, line width=\lineWidth, solid}}
\tikzset{RPEover/.style={RPE, mark=x, mark size=\markSize}}
\tikzset{RPEhist/.style={RPE, line width=\histWidth, solid, fill, ybar}}
\tikzset{GMM/.style={mark options={solid}, color=TUMBeamerRed, line width=\lineWidth, dashed}}
\tikzset{GMMover/.style={GMM, mark=triangle, mark size=\markSize}}
\tikzset{GMMhist/.style={GMM, line width=\histWidth, solid, fill, ybar}}
\tikzset{VAE/.style={mark options={solid}, color=ourdarkgreen, line width=\lineWidth, dashed}}
\tikzset{VAEover/.style={VAE, mark=pentagon, mark size=\markSize}}
\tikzset{VAEhist/.style={VAE, line width=\histWidth, solid, fill, ybar}}
\tikzset{GAN/.style={mark options={solid}, color=TUMBeamerOrange, line width=\lineWidth, dashed}}
\tikzset{GANover/.style={GAN, mark=o, mark size=\markSize}}
\tikzset{GANhist/.style={GAN, line width=\histWidth, solid, fill, ybar}}
\tikzset{DM/.style={mark options={solid}, color=mylila, line width=\lineWidth, dashed}}
\tikzset{DMover/.style={DM, mark=star, mark size=\markSize}}
\tikzset{DMhist/.style={DM, line width=\histWidth, solid, fill, ybar}}
\tikzset{scov/.style={mark options={solid}, color=TUMMediumGray, line width=\lineWidth, dashed}}
\tikzset{scovover/.style={scov, mark=|, mark size=\markSize}}
\tikzset{scovhist/.style={scov, line width=\histWidth, solid, fill, ybar}}

\newcommand{\legrpe}{RPE}
\newcommand{\leggmm}{GMM}
\newcommand{\legvae}{VAE}
\newcommand{\leggan}{WGAN}
\newcommand{\legdm}{DM}
\newcommand{\legscov}{scov}

\newcommand{\Nv}{{N_\text{v}}}
\newcommand{\Nh}{{N_\text{h}}}

\newcommand{\NL}{{N_\text{L}}}
\newcommand{\TR}{{T_\text{r}}}
\newcommand{\TV}{{T_\text{v}}}
\newcommand{\TE}{{T_\text{e}}}
\newcommand{\cnd}{{\,|\,}}

\newcommand{\herm}{^{\mkern0.5mu\mathrm{H}}}

\newcommand{\vmu}{{\bm{\mu}}}
\newcommand{\vsig}{{\bm{\sigma}}}
\newcommand{\veps}{{\bm{\varepsilon}}}
\newcommand{\vtheta}{{\bm{\theta}}}
\newcommand{\vphi}{{\bm{\phi}}}

\newcommand{\vc}{{\bm{c}}}

\newcommand{\vh}{{\bm{h}}}

\newcommand{\vn}{{\bm{n}}}

\newcommand{\vp}{{\bm{p}}}
\newcommand{\vq}{{\bm{q}}}

\newcommand{\vy}{{\bm{y}}}
\newcommand{\vz}{{\bm{z}}}

\newcommand{\mc}{{\bm{C}}}

\newcommand{\mf}{{\bm{F}}}

\newcommand{\mq}{{\bm{Q}}}

\newcommand{\RR}{{\mathbb{R}}}
\newcommand{\CC}{{\mathbb{C}}}
\newcommand{\diff}{{\mathrm{d}}}

\newcommand{\mmdhat}{\widehat{\textrm{MMD}}^2}

\DeclareMathOperator{\E}{E}

\DeclareMathOperator{\diag}{diag}
\DeclareMathOperator{\KL}{D_{KL}}

\DeclareMathOperator*{\argmax}{arg\,max}

\DeclareMathOperator{\eye}{\mathbf{I}}

\newacronym{tdd}{TDD}{time division duplex}
\newacronym{fdd}{FDD}{frequency division duplex}
\newacronym{lmmse}{LMMSE}{linear minimum mean squared error}
\newacronym{mse}{MSE}{mean squared error}
\newacronym{mmse}{MMSE}{minimum mean squared error}
\newacronym{nmse}{NMSE}{normalized mean squared error}
\newacronym{mimo}{MIMO}{multiple-input multiple-output}
\newacronym{simo}{SIMO}{single-input multiple-output}
\newacronym{miso}{MISO}{multiple-input single-output}
\newacronym{siso}{SISO}{single-input single-output}
\newacronym{deep}{DL}{deep learning}
\newacronym{ofdm}{OFDM}{orthogonal frequency division multiplexing}
\newacronym{csi}{CSI}{channel state information}
\newacronym{ula}{ULA}{uniform linear array}
\newacronym{ura}{URA}{uniform rectangular array}
\newacronym{dft}{DFT}{discrete fourier transform}
\newacronym{bs}{BS}{base station}
\newacronym{mt}{MT}{mobile terminal}
\newacronym{ae}{AE}{autoencoder}
\newacronym{ml}{ML}{machine learning}
\newacronym{dl}{DL}{deep learning}
\newacronym{doa}{DoA}{direction of arrival}
\newacronym{dod}{DoD}{direction of departure}
\newacronym{kl}{KL}{Kullback-Leibler}
\newacronym{elbo}{ELBO}{evidence lower bound}
\newacronym{iid}{i.i.d.}{independent and identically distributed}
\newacronym{fc}{FC}{fully connected}
\newacronym{nn}{NN}{neural network}
\newacronym{dnn}{NN}{neural network}
\newacronym{cnn}{CNN}{convolutional neural network}
\newacronym{ls}{LS}{least squares}
\newacronym{snr}{SNR}{signal-to-noise ratio}
\newacronym{se}{SE}{spectral efficiency}
\newacronym{ce}{CE}{channel estimation}
\newacronym{ul}{UL}{uplink}
\newacronym{mpc}{MPC}{multipath component}
\newacronym{rt}{RT}{ray-tracing}
\newacronym{mmd}{MMD}{maximum mean discrepancy}
\newacronym{tvd}{TVD}{total variation distance}
\newacronym{bleu}{BLEU}{bilingual evaluation understudy}
\newacronym{fid}{FID}{Fréchet inception distance}
\newacronym{cdf}{CDF}{cumulative distribution function}
\newacronym{pdf}{PDF}{probability density function}
\newacronym{tpr}{TPR}{true positive rate}
\newacronym{ccm}{CCM}{channel covariance matrix}
\newacronym{cg}{CG}{conditionally Gaussian}
\newacronym{3gpp}{3GPP}{3rd Generation Partnership Project}
\newacronym{vae}{VAE}{variational autoencoder}
\newacronym{vi}{VI}{variational inference}
\newacronym{cc}{CC}{convolutional channel}
\newacronym{cl}{CL}{convolutional layer}
\newacronym{ll}{LL}{linear layer}
\newacronym{rl}{RL}{reshaping layer}
\newacronym{bn}{BN}{batch normalization}
\newacronym{cs}{CS}{compressed sensing}
\newacronym{amp}{AMP}{approximate message passing}
\newacronym{gmm}{GMM}{Gaussian mixture model}
\newacronym{nf}{NF}{normalizing flow}
\newacronym{gan}{GAN}{generative adversarial network}
\newacronym{dm}{DM}{diffusion model}
\newacronym{vdm}{VDM}{variational diffusion model}
\newacronym{cme}{CME}{conditional mean estimator}
\newacronym{los}{LOS}{line of sight}
\newacronym{nlos}{NLOS}{non-line of sight}
\newacronym{rv}{RV}{random variable}
\newacronym{gm}{GM}{generative model}
\newacronym{psd}{PSD}{positive semi-definite}
\newacronym{wss}{WSS}{wide-sense stationary}
\newacronym{map}{MAP}{maximum a posteriori}
\newacronym{em}{EM}{expectation-maximization}
\newacronym{wd}{W1D}{Wasserstein-1 distance}
\newacronym{wgan}{WGAN}{Wasserstein GAN}
\newacronym{gp}{GP}{gradient penalty}
\newacronym{rpe}{RPE}{radio propagation environment}

%% file: content/abs.tex
Generative models are typically evaluated by direct inspection of their generated samples, e.g., by visual inspection in the case of images. Further evaluation metrics like the Fréchet inception distance or maximum mean discrepancy are intricate to interpret and lack physical motivation. These observations make evaluating generative models in the wireless PHY layer non-trivial. This work establishes a framework consisting of evaluation metrics and methods for generative models applied to the wireless PHY layer. The proposed metrics and methods are motivated by wireless applications, facilitating interpretation and understandability for the wireless community. In particular, we propose a spectral efficiency analysis for validating the generated channel norms and a codebook fingerprinting method to validate the generated channel directions. Moreover, we propose an application cross-check to evaluate the generative model's samples for training machine learning-based models in relevant downstream tasks. Our analysis is based on real-world measurement data and includes the Gaussian mixture model, variational autoencoder, diffusion model, and generative adversarial network as generative models. Our results under a fair comparison in terms of model architecture indicate that solely relying on metrics like the maximum mean discrepancy produces insufficient evaluation outcomes. In contrast, the proposed metrics and methods exhibit consistent and explainable behavior.

%% file: content/introduction.tex
\section{Introduction}
\label{sec:intro}

\IEEEPARstart{L}{earning} environment-specific features by training \acp{nn} on data from a particular \ac{rpe} recently emerged as a disruptive paradigm for the development of future wireless systems~\cite{Yu2022}. 
The development of novel wireless algorithms based on \ac{ml} has reached a mature stage and has also found its way into standardization~\cite{3GPP2023}. 
Methods such as CsiNet exemplarily stand for the success of \ac{ml}-based wireless systems, enabling an efficient feedback design for \ac{fdd} systems~\cite{Wen2018}. 
Other application examples from the wireless PHY layer include channel estimation~\cite{Ye2018, Neumann2018, Soltani2019} or precoding~\cite{Shi2021}.

Generative models are among the most popular \ac{ml}-based techniques as they enable learning data distributions based on samples~\cite{Ruthotto2021}. 
After successful training, the generative model allows for data likelihood evaluation and the creation of entirely new samples that follow the captured distribution. 
Well-known generative models are: the \ac{gmm}~\cite{Bishop2006}, \ac{vae}~\cite{Rezende2014, Kingma2014}, \ac{gan}~\cite{Goodfellow2020}, and the \ac{dm}~\cite{Ho2020}, the latter currently marking the state-of-the-art generation concept in image and speech processing.
Most of these models are originally proposed to generate natural signals such as images or speech. Still, they generally work well for arbitrary data and can also be utilized for inference tasks. 
Thus, generative models also find their way into the wireless literature due to their ability to model complex data relationships. 
For example, a generative model can build the foundation of an improved network management framework~\cite{Liu2024}. 
A generative model can also be used to design a semantics-aware vehicular network~\cite{Zhang2024}. 
For the wireless PHY layer, generative models demonstrate great potential in typical applications as well. 
Channel estimation, categorized as an inverse problem, is a domain where generative models are heavily applied to achieve remarkable estimation quality~\cite{Koller2022, Baur2023, Baur2024, Fesl2024, Balevi2021, Balevi2021a, Doshi2022, Arvinte2023, Fesl2024dm}. 
Moreover, generative models are also promising tools for model order selection~\cite{Baur2023modelorder}, feedback and precoder design~\cite{Turan2024}, channel prediction~\cite{Turan2024wsa}, or channel modeling~\cite{Xia2022a, Xiao2022, Euchner2024}. 

Almost all of the generative model-based wireless methods above are utilized for inference tasks after their training and not for generating samples. 
Most methods' performance is determined by assessing how well the generative model-based approach performs in a follow-up task, e.g., channel estimation. 
Indeed, in the wireless literature, there is a shortcoming concerning the direct evaluation of a generative model based on its generated samples. 
This observation contrasts other research disciplines, like image processing, where a fundamental aspect of a generative model's performance evaluation involves the quality of a generated sample, e.g., by visually inspecting a generated image. 
Therein, metrics like the \ac{fid}~\cite{Heusel2017} or \ac{mmd}~\cite{Gretton2012} are two well-known metrics to evaluate a generative model directly on its generated samples. 
Obviously, there is no such thing as visual quality for wireless data, necessitating alternative means of evaluation. 
A step in this direction is done by the channel modeling methods in~\cite{Xia2022a, Xiao2022, Euchner2024} as they evaluate specific channel parameters like power or delay spread based on the generated channels and compare them with the original data.
In~\cite{Xiao2022}, the authors go further and train an \ac{ae}-based method for channel compression on generated samples. 
They compare the \ac{ae}'s performance with an equivalent architecture \ac{ae} that was trained on the (synthetic) \ac{rpe} data to quantify the distributional shift between the \ac{rpe} and generative model channel distributions.

However, the task-specific evaluation for generative models in channel modeling only takes into account the methods proposed in the respective works and does not consider a broad evaluation of other generative models since this is not within the scope of their work. 
Moreover, the task-specific measures rely on knowing existing channel model parameters, which need to be estimated for measured data from a particular \ac{rpe}, limiting the validity of a channel parameter comparison. 
As a result, there is no clear guideline for assessing the quality of a generated channel with an established metric. 
Therefore, we propose a framework consisting of several evaluation metrics and methods for generative models in this work that are specifically suited for wireless channel data. 
We perform an extensive usage of the proposed metrics by applying them to the most popular generative models for PHY layer algorithms, i.e., the \ac{gmm}, \ac{vae}, \ac{gan}, and \ac{dm}. 
Moreover, the evaluation is done for real-world measurement data. 
The proposed evaluation metrics are motivated by PHY layer applications and have an interpretable character. 
In particular, the first evaluation metric we propose compares the achieved \ac{se} from the \ac{rpe} and generated samples marking the information-theoretic transmission bound. 
Second, we introduce a so-called \textit{codebook fingerprint}, where each dataset has a distinct fingerprint in terms of a codebook typically used in limited feedback schemes~\cite{Love2008}. 
The codebook fingerprint allows for an immediate distinction between channel datasets by visualizing them as discrete probability distributions as a consequence of normalizing empirically obtained histograms and utilizing the \ac{tvd} as a distance measure. 
Third, we propose an application cross-check where a model is trained on \ac{rpe} data, and another model is trained on generated data from a generative model that was trained on \ac{rpe} data. 
Afterward, both models are evaluated for the same task to determine how well the generative model can convey its channel distribution knowledge. 
Accordingly, the application cross-check provides a general framework where the specific task of interest can be customized. 
We summarize the primary \textit{contributions} as follows:
\begin{itemize}
    \item We give an overview of the state-of-the-art generative models for wireless applications and evaluate all of them on the proposed evaluation metrics in a fair comparison in terms of the models' architectures based on real-world measurements.
    \item We propose an \ac{se} analysis and codebook fingerprinting as evaluation routines, allowing us to assess whether the generative model can capture two important channel properties for PHY layer algorithms: the overall channel power and directional information.
    \item We introduce a general framework for an application cross-check. The cross-check enables us to determine if a generative model can be utilized in place of the true \ac{rpe} data for training \ac{ml}-based PHY layer methods and generate authentic channels.
\end{itemize}

To the best of our knowledge, our work is the first to establish general evaluation routines for generative models tailored specifically toward wireless PHY layer applications. 
The routines are easy to implement and interpret yet expressive enough to yield a reliable performance measure. 
Moreover, our extensive simulations show that, overall, the \ac{gmm} achieves the best generative quality followed by the \ac{vae}. 
A possible explanation for this behavior is that a \ac{gmm} or \ac{vae} can explicitly incorporate structural knowledge, such as a Toeplitz covariance matrix for uniformly structured arrays~\cite{Fesl2022}. 
The \ac{gan} or \ac{dm} need to learn this from scratch, an obvious disadvantage. 
However, we note that this outcome depends on the considered system configuration and that other system layouts may result in different rankings, requiring a system-specific evaluation. 

This paper is organized as follows. 
In Section~\ref{sec:prelim}, we give an overview of the state-of-the-art generative models for wireless applications, i.e., the \ac{gmm}, \ac{vae}, \ac{dm}, and \ac{gan}.
In Section~\ref{sec:eval_metrics}, we introduce the proposed evaluation metrics and methods for generative models in the wireless PHY layer. 
In Section~\ref{sec:results}, we describe the conducted measurement campaign, followed by the simulation results. 
We conclude this work in Section~\ref{sec:conclusion}.
In Appendix~\ref{app:implementation}, we list implementation details.

%% file: content/preliminaries.tex
\section{Generative Model Preliminaries}
\label{sec:prelim}

The problem in contemporary generative modeling is to learn a data distribution based on samples to create novel samples that follow the same distribution. 
More precisely, a generative model utilizes a set of training data samples $\{\vh_i\}_{i=1}^{\TR}\subset\CC^N$ with $\vh\sim p(\vh)$ to learn $p(\vh)$ with a parametric approximation $p_\vtheta(\vh)$.
The vector $\vtheta$ contains all the model parameters to specify the learned distribution, e.g., the \ac{nn} weights. 
One of the simplest methods to obtain $\vtheta$ is to assume a Gaussian distribution for $\vh$, i.e., $\vh\sim\mathcal{N}_{\CC}(\vmu_s, \mc_s)$ with the sample mean $\vmu_s=\frac{1}{\TR}\sum_{i=1}^\TR \vh_i$ and sample covariance $\mc_s=\frac{1}{\TR}\sum_{i=1}^\TR (\vh_i-\vmu_s) (\vh_i-\vmu_s)\herm$ such that $\vtheta=\{\vmu_s,\mc_s\}$.
We will refer to it as \textit{scov} in this work. 
An apparent disadvantage of the scov model is its limited expressiveness due to the single mean and covariance. 
An elegant way to obtain a more expressive generative model is to let the Gaussian distribution only hold conditionally, enforcing a \ac{cg} model. 
In particular, the \ac{cg} likelihood model becomes
\begin{equation}
    \vh\cnd\vz \sim p_\vtheta(\vh\cnd\vz) = \mathcal{N}_{\CC}(\vh; \vmu_\vtheta(\vz), \mc_\vtheta(\vz))
    \label{eq:cg}
\end{equation}
with the latent vector $\vz\in\RR^\NL$ following a prior $p(\vz)$.
To acquire $\vtheta$, a likelihood optimization-based approach is typically adopted to maximize $p_\vtheta(\vh)$.


\subsection{Gaussian Mixture Model}
\label{subsec:gmm}

\begin{algorithm}[t]
    \caption{Sampling from a GMM.}
    \label{alg:sampling_gmm}
    \begin{algorithmic}[1]
        \Require \ac{gmm} with parameters $\vtheta=\{\pi_k,\vmu_k,\mc_k\}_{k=1}^K$, number of samples to generate $M$.
        \For{$m=1,\ldots,M$}
            \State $k \sim \mathrm{Cat}([\pi_1,\ldots,\pi_K])$
            \State $\bm \varepsilon \sim \mathcal{N}_{\mathbb{C}}(\bm 0, \mathbf{I})$
            \State $\tilde{\vh}_m \gets \vmu_k + \mc_k^{1/2}\, \bm\varepsilon$
        \EndFor
        \State \Return \ac{gmm} generated dataset $\{\tilde\vh_m\}_{m=1}^M$.
    \end{algorithmic}
\end{algorithm}

When the condition is a discrete \ac{rv}, the number of Gaussian distributions in~\eqref{eq:cg} becomes finite.
Therefore, the corresponding likelihood function reads as
\begin{equation}
    p_\vtheta(\vh) = \sum_{k=1}^K \pi_k\, \mathcal{N}_{\CC}(\vh; \vmu_k,\mc_k)
    \label{eq:gmm_h}
\end{equation}
with the \textit{mixing coefficients} $\pi_k$ such that $\sum_{k=1}^K \pi_k = 1$ and $\vtheta=\{\pi_k,\vmu_k,\mc_k\}_{k=1}^K$ describing a \ac{gmm}~\cite{Bishop2006}.
The mean $\vmu_k$ and covariance $\mc_k$ belong to the $k$-th of in total $K$ Gaussian components. 
The $\pi_k$ address the components' weighting.
Due to the discrete-valued latent space, the posterior or \textit{responsibility} can be calculated in closed-form as
\begin{equation}
    p_\vtheta(k\cnd\vh) = \frac{\pi_k\, \mathcal{N}_{\CC}(\vh; \vmu_k,\mc_k)}{\sum_{\ell=1}^K \pi_\ell\, \mathcal{N}_{\CC}(\vh; \vmu_\ell,\mc_\ell)}.
\end{equation}
With the training data samples, a \ac{gmm} can be fitted with the \ac{em} algorithm to maximize $p_\vtheta(\vh)$; cf.~\cite{Bishop2006} for details. 
The sampling process from a fitted \ac{gmm} works as follows. 
First, a discrete realization from a categorical distribution with probabilities according to the mixing coefficients $\{\pi_i\}_{i=1}^K$ is drawn, determining the \ac{gmm} component. 
Second, the selected component is sampled as it is typically done for a Gaussian distribution.
Algorithm~\ref{alg:sampling_gmm} summarizes the procedure for in total $M$ samples.


\subsection{Variational Autoencoder}
\label{subsec:vae}

Although the \ac{gmm} is a universal approximator when the number of components grows to infinity~\cite{NgNgChMc20}, the \ac{gmm} fitting process with the \ac{em} algorithm and the responsibility evaluation become problematic with both $K$ and $N$ becoming large.
The \ac{vae} circumvents this problem by adopting~\eqref{eq:cg} with a continuous condition and parameterizing the mapping from $\vz$ to $\{\vmu_\vtheta(\vz), \mc_\vtheta(\vz)\}$ as a \ac{nn}, making the \ac{vae} more suitable for high-dimensional data.
However, the \ac{vae}-parameterized posterior 
\begin{equation}
    p_\vtheta(\vz\cnd\vh) = \frac{p_\vtheta(\vh\cnd\vz) p(\vz)}{\int p_\vtheta(\vh\cnd\vz) p(\vz) \diff\vz}
    \label{eq:posterior_vae}
\end{equation}
turns out to be intractable due to the continuous latent space, necessitating an approximate Bayesian technique for the model parameter optimization.
To this end, the likelihood is decomposed as~\cite{Baur2024}
\begin{equation}
    \log p_\vtheta(\vh) = \mathcal{L}_{\vtheta,\vphi}(\vh) + \KL(q_\vphi(\vz\cnd\vh)\,\|\,p_\vtheta(\vz\,|\,\vh))
    \label{eq:log-like}
\end{equation}
with the \ac{elbo}
\begin{equation}
    \mathcal{L}_{\vtheta,\vphi}(\vh) = \E_{q_{\vphi}} \left[\log p_{\vtheta}(\vh\cnd\vz)\right] - \KL(q_{\vphi}(\vz\cnd\vh)\,\|\,p(\vz))
    \label{eq:vae}
\end{equation}
and the non-negative \ac{kl} divergence 
\begin{equation}
    \KL(q_\vphi(\vz\cnd\vh)\,\|\,p_\vtheta(\vz\cnd\vh)) = \E_{q_\vphi}\left[ \log \left( \frac{q_\vphi(\vz\cnd\vh)}{p_\vtheta(\vz\cnd\vh)} \right) \right].
    \label{eq:elbo-gap}
\end{equation}
We write $\E_{q_\vphi(\vz|\vh)}[\cdot] = \E_{q_\vphi}[\cdot]$ for notational brevity.
In~\eqref{eq:log-like}, $q_\vphi$ is introduced to approximate the posterior~\eqref{eq:posterior_vae}, becoming apparent by noticing that an \ac{elbo} maximization not only maximizes the likelihood but also minimizes~\eqref{eq:elbo-gap}.

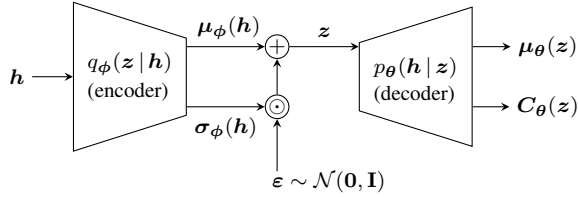
\begin{figure}[t]
    \centering
    \input{tikz_figures/vae}
    \caption{Structure of a VAE with \ac{cg} distributions for $q_{\vphi}(\vz\cnd\vh)$ and $p_{\vtheta}(\vh\cnd\vz)$. The encoder and decoder each represent a \ac{nn}.}
    \label{fig:vae}
\end{figure}

Despite $p_\vtheta(\vz\cnd\vh)$ being non-Gaussian according to~\eqref{eq:posterior_vae}, its approximation is commonly defined as
\begin{equation}
    q_\vphi(\vz\cnd\vh) = \mathcal{N}(\vz; \vmu_\vphi(\vh),\diag(\vsig^2_\vphi(\vh)))
    \label{eq:q_phi}
\end{equation}
and the prior $p(\vz)=\mathcal{N}(\bm{0}, \mathbf{I})$ due to optimization purposes involving their simplicity to be sampled from.
Moreover, the \ac{vae} also implements $q_\vphi(\vz\cnd\vh)$ via a \ac{nn}.
Fig.~\ref{fig:vae} illustrates the described \ac{vae}'s working principle. 
The encoder receives a channel $\vh$ as input and maps it to the first two moments of $q_\vphi(\vz\cnd\vh)$, i.e., $\vmu_\vphi(\vh)$ and $\vsig_\vphi(\vh)$.
Afterward, the \textit{reparameterization trick} is applied to obtain $\vz = \vmu_\vphi(\vh)+\vsig_\vphi(\vh)\,\odot\,\veps$. 
The sample $\vz$ is the decoder input, being mapped to $\vmu_\vtheta(\vz)$ and $\mc_\vtheta(\vz)$ representing the first two moments of $p_{\vtheta}(\vh\cnd\vz)$.
Furthermore, the \ac{cg} distributions allow for analytic expressions in the \ac{elbo}~\cite{Baur2024}. 
Finally, the \ac{vae}'s sampling process is summarized in Algorithm~\ref{alg:sampling_vae}. 
In step two, the latent vector is drawn from a standard Gaussian distribution. 
The \ac{vae} decoder yields the first two moments of $p_\vtheta(\vh\cnd\vz)$, which are utilized to generate a channel sample in the fourth step.

\begin{algorithm}[t]
    \caption{Sampling from a VAE.}
    \label{alg:sampling_vae}
    \begin{algorithmic}[1]
        \Require \ac{vae} decoder \ac{nn} with parameters $\vtheta$, number of samples to generate $M$
        \For{$m=1,\ldots,M$}
            \State $ \vz \sim \mathcal{N}(\bm 0, \mathbf{I})$
            \State $ \bm \varepsilon \sim \mathcal{N}_{\mathbb{C}}(\bm 0, \mathbf{I})$
            \State $\tilde\vh_m \gets \vmu_\vtheta(\vz) + \mc^{1/2}_\vtheta(\vz)\, \bm \varepsilon $
        \EndFor
        \State \Return \ac{vae} generated dataset $\{\tilde\vh_m\}_{m=1}^M$.
    \end{algorithmic}
\end{algorithm}

\subsection{Diffusion Model}
\label{subsec:dm}

\begin{figure}[t]
\centering
\input{tikz_figures/dm}
\caption{Markov chain of the DM involving the forward process with $q(\vh_{t}\cnd \vh_{t-1})$ and the approximated reverse process with $p_\vtheta(\vh_{t-1}\cnd \vh_{t})$.}
\label{fig:dm_chain}
\end{figure}
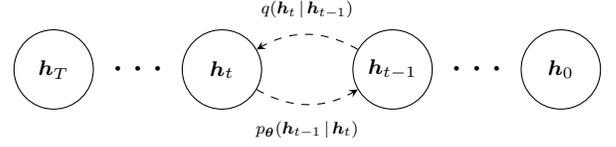

The principal concept of the \ac{dm} is similar to the \ac{vae} in the sense they share the \ac{elbo} as an optimization objective. 
Yet, the \ac{dm}'s internal functioning significantly differs from the \ac{vae}. 
The diffusion process involves a repeated addition of noise such that a clean data sample $\vh_0$ becomes pure noise $\vh_T$ at the end of the \textit{forward process}, which goes from $0$ to $T$ in Fig.~\ref{fig:dm_chain} and defines a Markov chain~\cite{Ho2020}.
In particular,
\begin{equation}
    \vh_{t} = \sqrt{\alpha_t} \vh_{t-1} + \sqrt{1 - \alpha_t} \bm{\varepsilon}, \quad \bm{\varepsilon} \sim \mathcal{N}(\bm 0, \mathbf{I})
    \label{eq:dm_forward}
\end{equation}
so
\begin{equation}
    q(\vh_{t}\cnd \vh_{t-1}) = \mathcal{N}(\vh_{t}; \sqrt{\alpha_t} \vh_{t-1}, (1 - \alpha_t) \mathbf{I})
    \label{eq:dm_prob_forward}
\end{equation}
with $t = 1,\ldots,T$.
The time-dependent hyperparameter $\alpha_t$ controls the noise variance in every step and is either learnable or follows a fixed schedule.

To obtain samples with the \ac{dm}, the \textit{reverse process} represented by $p(\vh_{t-1}\cnd \vh_{t})$ is utilized, which is analytically inaccessible, necessitating an approximation $p_\vtheta(\vh_{t-1}\cnd \vh_{t})$.
For determining the structure of $p_\vtheta(\vh_{t-1}\cnd \vh_{t})$, it is beneficial to investigate the \ac{dm}'s \ac{elbo}: 
\begin{align}
     & \E_{q(\vh_1\cnd\vh_0)}[\log p_\vtheta(\vh_0\cnd\vh_1)] - \KL(q(\vh_T\cnd\vh_0) \,\|\, p(\vh_T)) \nonumber \\
     - & \sum_{t=2}^T \E_{q(\vh_t\cnd\vh_0)}[\KL(q(\vh_{t-1}\cnd\vh_t,\vh_0) \,\|\, p_\vtheta(\vh_{t-1}\cnd \vh_{t}))].
\end{align}
The third term is most relevant for the \ac{dm}'s training as it involves the complete Markov chain except for one step, while the first term is usually neglected, and the second term is assumed to be zero.
Consequently, $p_\vtheta(\vh_{t-1}\cnd \vh_{t})$ should be structurally equal to $q(\vh_{t-1}\cnd\vh_t,\vh_0) = \mathcal{N}(\vh_{t-1}; \vmu_q(\vh_t,\vh_0), \sigma_t^2\mathbf{I})$ for an \ac{elbo} maximization, resulting in
\begin{equation}
    p_\vtheta(\vh_{t-1}\cnd \vh_{t}) = \mathcal{N}(\vh_{t-1}; \vmu_{\vtheta}(\vh_t, t), \sigma^2_t\eye).
    \label{eq:dm_prob_reverse}
\end{equation}
The mean $\vmu_{\vtheta}(\vh_t, t)$ is learned by a \ac{nn} and 
\begin{equation}
    \sigma_t^2 = \frac{(1-\alpha_t)(1 - \bar{\alpha}_{t-1})}{1 - \bar{\alpha}_t}.
\end{equation}
represents the time-dependent variance with $\bar{\alpha}_t = \prod_{i=1}^t \alpha_i$.

For a more detailed \ac{dm} introduction, we refer the reader to~\cite{Luo2022}. 
Without loss of generality, we have adopted real-valued distributions in this section. 
In this case, a channel's real and imaginary parts would be stacked to yield a real-valued $\vh_0.$
We describe the \ac{dm}'s sampling procedure in Algorithm~\ref{alg:sampling_dm}. 
The concept is to sample from a standard Gaussian distribution and then repeatedly sample from the reverse process with $p_\vtheta(\vh_{t-1}\cnd \vh_{t})$ until a noise-free sample is reached. 
In step seven, we indicate that the real-valued sample must be converted to the complex domain at the end.

\begin{algorithm}[t]
    \caption{Sampling from a \ac{dm}.}
    \label{alg:sampling_dm}
    \begin{algorithmic}[1]
        \Require \ac{dm} with parameters $\vtheta$, number of samples to generate $M$.
        \For{$m=1,\ldots,M$}
            \State $ \vh \sim \mathcal{N}(\bm 0, \mathbf{I})$
            \For{$t=T,\ldots,1$}
                \State $ \bm \varepsilon \sim \mathcal{N}(\bm 0, \mathbf{I})$
                \State $ \vh \gets \vmu_{\vtheta}(\vh, t) + \sigma_t\, \bm\varepsilon $
            \EndFor
            \State $ \tilde\vh_m \gets \mathrm{real2complex}(\vh) $
        \EndFor
        \State \Return \ac{dm} generated dataset $\{\tilde\vh_m\}_{m=1}^M$.
    \end{algorithmic}
\end{algorithm}

\subsection{Generative Adversarial Network}
\label{subsec:gan}

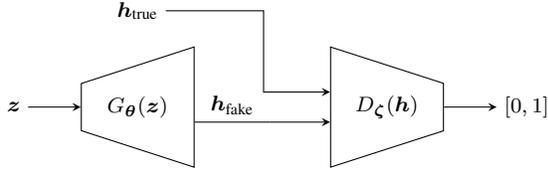
\begin{figure}[t]
    \centering
    \input{tikz_figures/gan}
    \caption{Illustration of a GAN with generator $G_\vtheta(\vz)$ and discriminator $D_{\bm{\zeta}}(\vh)$ each representing a \ac{nn}.}
    \label{fig:gan}
\end{figure}

The last generative model we discuss in this section is the \ac{gan}~\cite{Goodfellow2020}.
A \ac{gan} consists of a generator $G_\vtheta(\vz)$ and a discriminator $D_{\bm{\zeta}}(\vh)$, illustrated in Fig.~\ref{fig:gan}, where $\vz$ is sampled from $\mathcal{N}(\bm{0},\mathbf{I})$.
The discriminator's task is to tell whether a sample stems from the true data distribution or the generator, while the generator mimics original data samples as well as possible. 
This competing behavior is reflected in the \ac{gan}'s minimax training objective. 
It is well-known that a \ac{gan} is intricate to train, often suffering from unstable training, mode collapse, or vanishing gradient. 
Many adaptions have been proposed in the literature to cope with these limitations, of which the \ac{wgan} is a prominent case~\cite{Arjovsky2017}.
The \ac{wgan}'s training strategy reads as
\begin{equation}
    \min_{G_\vtheta} \max_{D_{\bm\zeta}} \E_{p(\vh)}[D_{\bm\zeta}(\vh)] - \E_{p(\vz)}[D_{\bm\zeta}(G_\vtheta(\vz))]
\end{equation}
minimizing the \ac{wd}.
Additionally, a \ac{gp} can be incorporated into the \ac{wgan} to improve the generative performance~\cite{Gulrajani2017}. 
Since sampling from the \ac{wgan} only involves passing $\vz$ through the generator, we omit a separate sampling algorithm here.

%% file: tikz_figures/vae.tex
\begin{tikzpicture}[>=stealth]
	\def\NNsize{1.7cm}
	\def\NNwidth{1.5cm}
	\def\NNheight{1.6cm}
	\def\NNangle{70}
	\small{
	\node (input) at (0, 0) { $\bm{h}$};
	
	\node[trapezium, draw, align=center, trapezium stretches = true, minimum height=\NNwidth, minimum width=\NNheight, align=center, trapezium angle=\NNangle, rotate=-90] (NN1) at($ (input) + (1.5,0) $) {\small{ \rotatebox{90}{\hspace{-1mm} $q_{\vphi}(\bm{z}\cnd\bm{h})$}} \small\rotatebox{90}{(encoder)} };
	
	\node[circle, draw, inner sep = 0.01em] (add) at ($ (NN1.north) + (1.2,0.4) $) {$ + $};
	\node[circle, draw, inner sep = 0.01em] (multi) at ($ (NN1.north) + (1.2,-0.4) $) {$ \odot $};
	\node (eps) at ( $(multi) + (0.3, -1)$ ) {\hspace{2.4em} $\bm{\varepsilon} \sim \mathcal{N}(\bm{0},\mathbf{I}) $};
	
	\node[trapezium, draw, align=center, trapezium stretches = true, minimum height=\NNwidth, minimum width=\NNheight, align=center, trapezium angle=\NNangle, rotate=90](NN2) at ($ (NN1) + (3.8,0) $) {\small{\rotatebox{-90}{(decoder)}} \small{\rotatebox{-90}{$p_{\vtheta}(\bm{h}\cnd\bm{z})$}} };
	
	
	\node(output_mu) at ($ (NN2.south) + (1,0.4) $) { $\vmu_\vtheta(\vz)$};
	\node(output_cov) at ($ (NN2.south) + (1,-0.4) $) { $\mc_\vtheta(\vz)$};
	
	\draw[->] (input.east) -- (NN1.south) {};
	
	\draw[->] ($ (NN1.north) + (0,0.4) $) --node[midway, above=-.0em]{\hspace{-0.2mm} $\vmu_\vphi(\bm{h})$}    (add.west);
	\draw[->] ($ (NN1.north) + (0,-0.4) $) --node[midway, above=-1.9em]{ $\vsig_\vphi(\bm{h})$} (multi.west);
	\draw[->] (multi.north) -- (add.south);
	\draw[->] ($ (eps.north) - (0.3, 0.1) $)   -- (multi.south);
	\draw[->] (add.east) --node[midway, above=.1em]{ $\bm{z}$} ($(NN2.north) + (0,0.4)$);
	
	\draw[->] ($ (NN2.south) + (0,0.4) $) -- (output_mu.west);
	\draw[->] ($ (NN2.south) + (0,-0.4) $) -- (output_cov.west);
	}
\end{tikzpicture}

%% file: tikz_figures/dm.tex
\begin{tikzpicture}[->, >=stealth, node distance=1.0cm, minimum size=30pt]
    \small{
    \node[circle, draw] (h_T) {};
    \node at (h_T) {$\vh_{T}$};
    \node[] (gap) [right=0.1cm of h_T] {$\scalemath{2}{\cdots}$};
    \node[circle, draw] (h_t) [right=0cm of gap] {};
    \node at (h_t) {$\vh_{t}$};
    \node[circle, draw] (h_t-1) [right=1.2cm of h_t] {};
    \node at (h_t-1) {$\vh_{t-1}$};
    \node[] (gap2) [right=0.1cm of h_t-1] {$\scalemath{2}{\cdots}$};
    \node[circle, draw] (three) [right=0cm of gap2] {};
    \node at (three) {$\vh_0$};
        
    \path(h_t) edge [bend right, dashed] node [below=-0.2cm]{ $\scalemath{0.8}{p_\vtheta(\vh_{t-1}\cnd \vh_{t})}$} (h_t-1);
    \path(h_t-1) edge [bend right, dashed] node [above=-0.2cm]{ $\scalemath{0.8}{q(\vh_{t}\cnd \vh_{t-1})}$} (h_t);
    }
\end{tikzpicture}

%% file: tikz_figures/gan.tex
\begin{tikzpicture}[>=stealth]
	\def\NNwidth{1.5cm}
	\def\NNheight{1.6cm}
	\def\NNangle{70}
    \small{
    \node[align=center,](z) at (0,0) {$\vz$}; 
    \node[draw, trapezium, shape border rotate=90,  align=center, right= 0.7cm of z, trapezium stretches = true, minimum height=\NNwidth, minimum width=\NNheight, trapezium angle=\NNangle]
    (G) {$G_\vtheta(\vz)$};
    \node[align=center, above=0.5cm of G,](x){$\vh_{\textrm{true}}$};
    \node[draw, trapezium, shape border rotate=270, align=center, right= 1.8cm of G, trapezium stretches = true, minimum height=\NNwidth, minimum width=\NNheight, trapezium angle=\NNangle]
    (D) {$D_{\bm{\zeta}}(\vh)$};
    
    \node[right=0.7cm of D](out){$[0,1]$};
    
    \draw[->] (z.east) -- (G.west);
    
    \draw[] ([yshift=-0.2cm]G.east) -- ([xshift=1cm, yshift=-0.2cm]G.east) node [midway, above]{$\vh_{\textrm{fake}}$};
    \draw[->] ([xshift=1cm, yshift=-0.2cm]G.east) -- ([yshift=-0.2cm]D.west);
    
    \draw[->] (x.east) -- ([xshift=1.3cm]x.east) |- ([yshift=0.2cm]D.west);
    \draw[->] (D.east) -- (out.west);
    }
\end{tikzpicture}

%% file: content/evaluation_metrics.tex
\section{Evaluation Metrics and Methods}
\label{sec:eval_metrics}

Evaluation routines for generative models are mainly driven by applications from image or natural language processing. 
For example, a generative model for image generation is usually evaluated with the generated images' perceptual quality in combination with metrics like the \ac{mmd} or \ac{fid}. 
Related metrics exist for generated texts, e.g., \ac{bleu}~\cite{Vaswani2017}. 
While visual or auditory inspections are established evaluation routines for corresponding data, such things are problematic regarding wireless channels. 
Furthermore, metrics like the \ac{mmd} that can straightforwardly be calculated for channels are themselves a \ac{rv} without an intuitive meaning, being only suitable in direct comparison to other generated data. 
Moreover, it is unclear in which way an \ac{mmd} value can be transferred to the performance in a relevant application. 
The limited applicability of existing evaluation methods for generative models motivates the proposal of novel techniques specifically tailored toward wireless channels that provide a physical or application-dependent interpretation.


\subsection{Spectral Efficiency Analysis}
\label{subsec:se-analysis}

The quality of a channel realization significantly influences the achievable \ac{se} of a communications system, mainly driven by the channel norm. 
A generative model trained on channel data from a particular \ac{rpe} should be able to generate channels that produce similarly distributed \acp{se} as the original data.
To this end, we evaluate the \ac{se} expression
\begin{equation}
    r(\vh) = \log_2 \left ( 1 + \frac{\|\vh\|_2^2}{\sigma^2} \right)
    \label{eq:se}
\end{equation}
for the system model
\begin{equation}
    \vy = \vh + \vn, \quad \vn \sim \mathcal{N}_\CC(\bm 0, \sigma^2\mathbf{I}).
    \label{eq:system-model}
\end{equation}
In~\eqref{eq:se}, the channel is either sampled from the original \ac{rpe} or a generative model to reveal possible differences.
For a means of comparing different \ac{se} distributions, the empirical \ac{cdf} is an appropriate tool for visualization since the \ac{se} is one-dimensional. 
The \ac{wd} between the \ac{rpe} and a generated \ac{se} distribution, which is the absolute area between the respective \acp{cdf}, provides a quantitative measure to support the visual perception.


\subsection{Codebook Fingerprinting}

The \ac{se} analysis from the previous section primarily contrasts the \ac{rpe} and generated channels' norms. 
For a complete wireless channel comparison, not only the norm but also the channel direction should be taken into account. 
Channel directions are relevant for many applications, especially precoder design~\cite{Love2008}. 
In \ac{fdd} systems, the \ac{bs} and \ac{mt} typically share a codebook, and the \ac{mt} only sends the most aligned codebook entry back to the \ac{bs} for precoding instead of the complete estimated channel to save feedback overhead.
More precisely, the codebook $\mathcal{C} = \{\vc_1, \ldots, \vc_C\}$ representing $B=\log_2 C$ feedback bits is used to determine the feedback index
\begin{equation}
    j = \argmax_n |\vc\herm_n\, \tilde\vh |, \quad n=1,\ldots,C
    \label{eq:feedback}
\end{equation}
for a representative channel sample $\tilde\vh$.

In a real-world system, $\tilde\vh$ would be a channel estimate. 
However, we can also determine the \ac{rpe} channels' feedback indices and characterize the directions in the \ac{rpe} with the codebook. 
After proper normalization, the result is a discrete probability measure or histogram. 
Similarly, we can create a histogram for channels sampled from a generative model. 
Each histogram is representative of the respective channel distribution and yields a \textit{codebook fingerprint}. 
By comparing different codebook fingerprints with each other, we propose another means of evaluating a generative model for wireless data. 
Since comparing histograms by inspection is tedious, especially for large codebooks, we additionally compute the \ac{tvd} between the probability measures.
In particular, for the two probability measures $P$ and $Q$ representing the codebook fingerprints, the \ac{tvd} is given by
\begin{equation}
    \delta(P,Q) = \frac{1}{2} \sum_{n=1}^C |P(n) - Q(n)|.
    \label{eq:tvd}
\end{equation}
The \ac{tvd} has the beneficial property that it lies between zero and one, where zero attests to a perfect match.


\subsection{Application Cross-Check}
\label{subsec:cross-check}

As a third way of evaluating a generative model's quality, we want to investigate how well it can transfer its knowledge to other data-driven methods, e.g., for channel estimation or compression. 
Concerning knowledge transfer, we mean, what performance gap do we obtain when training a data-driven method on \ac{rpe} samples compared to training with generative model samples?

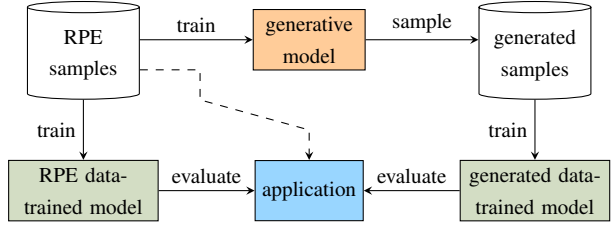
\begin{figure}[t]
    \centering
    \input{tikz_figures/cross_check}
    \caption{Illustration of the application cross-check.}
    \label{fig:cross-check}
\end{figure}

Fig.~\ref{fig:cross-check} shows an illustration of the adopted methodology. 
In the top row, a generative model utilizes \ac{rpe} samples for its training. 
Afterward, it generates new generative model samples that should follow the same distribution as the \ac{rpe} samples. 
In the bottom row on the left, a data-based model is trained on the \ac{rpe} samples for a specific application, while on the right, a model is trained on the generative model samples for the same application. 
The models in the bottom row do not necessarily need to be generative models. 
They can be any data-based model suitable for the selected application. 
Nevertheless, the models should be powerful enough to capture their training data distribution. 
The blue box in the middle symbolizes that the models are evaluated based on the same application with \ac{rpe} test samples. 
If the generative model properly learns the \ac{rpe} distribution, the \ac{rpe} data-trained model and generated data-trained model in the bottom row should perform similarly on the considered application since the distributional shift between the \ac{rpe} and generated samples becomes negligible.

As application blocks for Fig.~\ref{fig:cross-check}, we choose channel estimation and compression for the system model in~\eqref{eq:system-model}; two PHY layer applications where \ac{dl} methods are heavily used to improve the performance of a wireless communications system. 
Accurate channel estimates are vital to attaining the full potential of massive \ac{mimo} systems, so channel estimators that are powerful while exhibiting a low computational complexity are essential. 
What is more, channel compression is relevant for \ac{fdd} systems to reduce the feedback overhead after the channel has been estimated. 
To this end, we briefly describe the adopted channel estimators and the channel compression framework for the application cross-check in the following two subsections.

\subsubsection{Channel Estimators}
\label{subsubsec:channel_est}

One of the simplest channel estimators adopts a Gaussian prior and utilizes the sample mean $\vmu_s$ and sample covariance matrix $\mc_s$ to parameterize a \ac{lmmse} estimator.
In particular, the estimator reads as
\begin{equation}
    \hat{\vh}_\text{scov}(\vy) = \vmu_s + \mc_s (\mc_s + \sigma^2\mathbf{I})^{-1} (\vy - \vmu_s)
    \label{eq:est_lmmse}
\end{equation}
for the system model in~\eqref{eq:system-model}.

A more powerful estimator can be obtained by fitting a \ac{gmm} prior to the channel distribution and deriving an estimator aiming at minimizing the \ac{mse}~\cite{Koller2022}.
To compute the final estimate, this \ac{gmm}-based channel estimator requires a \ac{gmm} for $\vy$.
For this purpose, a \ac{gmm} is initially fitted for $\vh$ according to~\eqref{eq:gmm_h}.
Then, the \ac{gmm} is updated with the help of~\eqref{eq:system-model} to yield
\begin{equation}
    p_\vtheta(\vy) = \sum_{k=1}^K \pi_k\, \mathcal{N}_{\CC}(\vy; \vmu_k,\mc_k + \sigma^2\mathbf{I}).
    \label{eq:gmm_y}
\end{equation}
with the same $\vmu_k$ and $\mc_k$ as in~\eqref{eq:gmm_h} and $\sigma^2$ as additional parameter.
The resulting channel estimate adopting the \ac{gmm} prior is calculated as
\begin{equation}
    \hat{\vh}_\text{GMM}(\vy) = \sum_{k=1}^K p_\vtheta(k\cnd\vy) \left[\vmu_k + \mc_k (\mc_k + \sigma^2\mathbf{I})^{-1} (\vy - \vmu_k)\right]
    \label{eq:gmm_est}
\end{equation}
with 
\begin{equation}
    p_\vtheta(k\cnd\vy) = \frac{\pi_k\, \mathcal{N}_{\CC}(\vy; \vmu_k,\mc_k+\sigma^2\mathbf{I})}{\sum_{\ell=1}^K \pi_\ell\, \mathcal{N}_{\CC}(\vy; \vmu_\ell,\mc_\ell+\sigma^2\mathbf{I})}.
    \label{eq:resp_y}
\end{equation}
For details regarding the \ac{gmm}-based estimator derivation, we refer to~\cite{Koller2022}.

Another channel estimator leveraging a powerful generative prior is the \ac{vae}-based channel estimator~\cite{Baur2024}. 
This approach trains a \ac{vae} as in Fig.~\ref{fig:vae} with the distinction that the encoder receives $\vy$ as input since $\vh$ is inaccessible during the estimation phase. 
After its training, the \ac{vae} parameterizes $\vh\cnd\vz$ as \ac{cg}, which can be exploited to derive an estimator that minimizes the \ac{mse} adopting the \ac{vae} prior, cf.~\cite{Baur2024} for details.
The resulting \ac{vae}-based channel estimator reads as
\begin{equation}
    \hat{\vh}_\text{VAE}(\vy) = \vmu_\vtheta(\tilde\vz) + \mc_\vtheta(\tilde\vz) (\mc_\vtheta(\tilde\vz) + \sigma^2\mathbf{I})^{-1} (\vy - \vmu_\vtheta(\tilde\vz))
    \label{eq:vae_est}
\end{equation}
with $\tilde\vz=\vmu_\vphi(\vy)$ being the encoder mean and $\vmu_\vtheta(\tilde\vz)$ and $\mc_\vtheta(\tilde\vz)$ the corresponding decoder mean and covariance matrix, respectively, cf. Fig.~\ref{fig:vae}.

\subsubsection{Channel Compression}
\label{subsubsec:channel_compr}

\Ac{ae} models such as CsiNet~\cite{Wen2018} are the dominant way in the literature for the channel compression task and are essential parts of feedback schemes in \ac{fdd} systems~\cite{3GPP2023}. 
To reduce feedback overhead, the \ac{ae} compresses a channel estimate at the \ac{mt} by a specific factor $\rho$. 
This step is accomplished with the encoder of the \ac{ae}. 
Instead of feedbacking the complete channel estimate back to \ac{bs}, only the compressed version is fed back. 
At the \ac{bs} side, the channel estimate is reconstructed with the \ac{ae} decoder based on the compressed version. 
The \ac{ae} architecture is typically trained by minimizing the \ac{mse}.


\subsection{Further Evaluation Metrics}
\label{subsec:further-metrics}

In general, many evaluation metrics exist in the literature to evaluate the quality of generated samples~\cite{Bischoff2024}. 
Two of the most popular are the \ac{fid} and \ac{mmd}. 
Essentially, the metrics compute a dimensionality reduction that aims to preserve the statistical relations between the data. 
The \ac{mmd} uses kernels for this task~\cite{Gretton2012}, and the \ac{fid} leverages \acp{nn}~\cite{Heusel2017}.
We will focus on the \ac{mmd} in our further elaborations, but similar reasonings hold for the \ac{fid}.

Closed-form evaluation of the \ac{mmd} is impractical since only samples are provided for the original data without access to the generating distribution.
An unbiased \ac{mmd} estimate is thus typically computed as
\begin{equation}
    \mmdhat(p,q,\phi) = \frac{1}{n(n-1)} \sum_{i\neq j} g_{ij}
    \label{eq:mmd}
\end{equation}
with $g_{ij} = \phi(\vp_i,\vp_j) + \phi(\vq_i,\vq_j) - \phi(\vp_i,\vq_j) - \phi(\vq_i,\vp_j)$ based on the samples $\{\vp_i\}_{i=1}^L$ and $\{\vq_i\}_{i=1}^L$ from the original data distribution $p$ and its approximation $q$, respectively.
The kernel $\phi(\cdot,\cdot)$ is a conventional Gaussian kernel. 
In our case, $p$ will represent the \ac{rpe} and $q$ a generative model.

The problem with the \ac{mmd} is that it is a \ac{rv} itself with a possible value range between zero and infinity, preventing a direct interpretation. 
As we will demonstrate in Section~\ref{sec:results}, the \ac{mmd} values for different generative models will be comparable. 
Therefore, calculating the \ac{mmd} alone to evaluate a generative model would be insufficient, especially for wireless communications-related applications, and should only serve as an auxiliary metric. 
In contrast, the proposed evaluation metrics and methods from the previous sections are inherently interpretable and allow for an explainable generative model evaluation. 
In particular, the proposed evaluation metrics also perform a dimensionality reduction similar to the \ac{mmd}, but towards a well-known quantity like a codebook entry, making it more suitable for wireless communications-related problems.

%% file: tikz_figures/cross_check.tex
\begin{tikzpicture}[>=stealth]
\def\minHeight{1.3cm}
\def\textWidth{1.3cm}
\def\textWidthmodel{1.8cm}
\def\minSize{0.8cm}
\def\blockSep{1.5cm}
\def\blockBelow{0.8cm}
\def\yShift{0.2cm}
\def\connShift{0.4cm}
\small{

\node[draw, text width=\textWidth, minimum height=\minHeight, align=center, shape=cylinder, shape border rotate=90, shape aspect=.1, align=center,](RPE){RPE samples};

\node[draw, text width=\textWidth, fill=TUMBeamerOrange!40, minimum size=\minSize, align=center, yshift=\yShift, right= \blockSep of RPE](gm-orig) {generative model};

\node[draw, text width=\textWidth, minimum height=\minHeight, shape=cylinder, shape border rotate=90, shape aspect=.1, align=center, yshift=-\yShift, right=\blockSep of gm-orig](samples-gm) {generated samples};

\node[draw, fill=TUMBeamerGreen!40, minimum size=\minSize, text width=\textWidthmodel, align=center, below=\blockBelow of samples-gm](model-gm){generated data-trained model};

\node[draw, fill=TUMBeamerGreen!40, minimum size=\minSize, text width=\textWidthmodel, align=center, below= \blockBelow of RPE](model-rpe) {RPE data-trained model};

\node[draw, fill=TUMBeamerBlue!40, minimum size=\minSize, align=center](eval) at ($(model-rpe)!0.5!(model-gm)$) {application};

\draw[->] ([yshift=\yShift]RPE.east) -- (gm-orig.west) node[midway, above]{train};
\draw[->] (gm-orig.east) -- ([yshift=\yShift]samples-gm.west) node[midway, above]{sample};
\draw[->] (samples-gm.south) -- (model-gm.north) node[midway, left]{train};
\draw[->] (RPE.south) -- (model-rpe.north) node[midway, left]{train};
\draw[->] (model-rpe.east) -- (eval.west) node[midway, above]{evaluate};
\draw[->] (model-gm.west) -- (eval.east) node[midway, above]{evaluate};
\draw[->, dashed] ([yshift=-\yShift]RPE.east) -- ([yshift=-\yShift]$(RPE.east) + (0.8, 0)$) -- ([yshift=-\yShift]$(RPE.east) + (0.8, -\connShift)$) -- ([yshift=\connShift]eval.north) --  (eval.north);
}
\end{tikzpicture}

%% file: content/results.tex
\section{Evaluation on Real-World Measurement Data}
\label{sec:results}


\subsection{Measurement Campaign}
\label{subsec:meas_campaign}

\begin{figure}[t]
    \centering
    \begin{tikzpicture}
        \draw (0,0) node[below right] {\includegraphics[width=0.975\columnwidth]{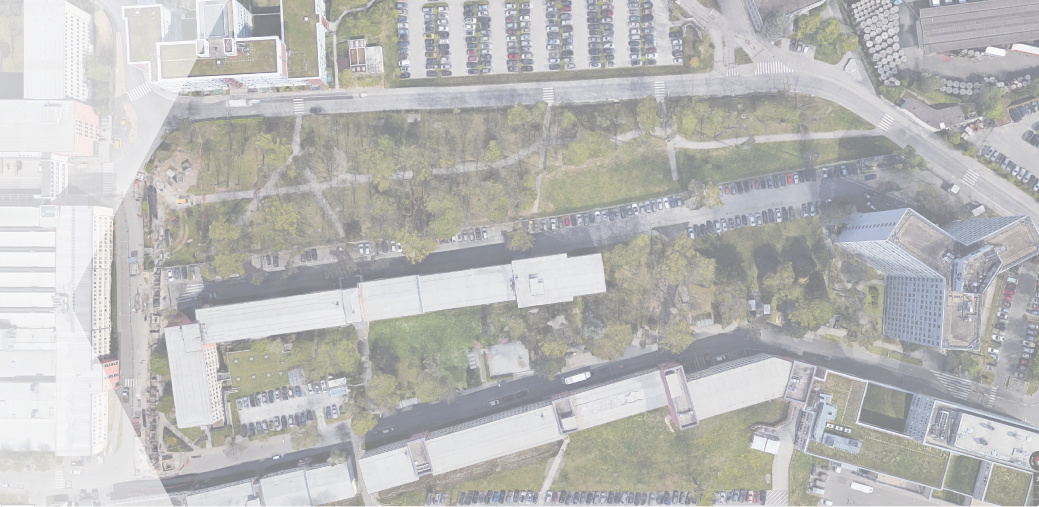}};
        \draw (0,0) node[below right] {\includegraphics[width=0.975\columnwidth]{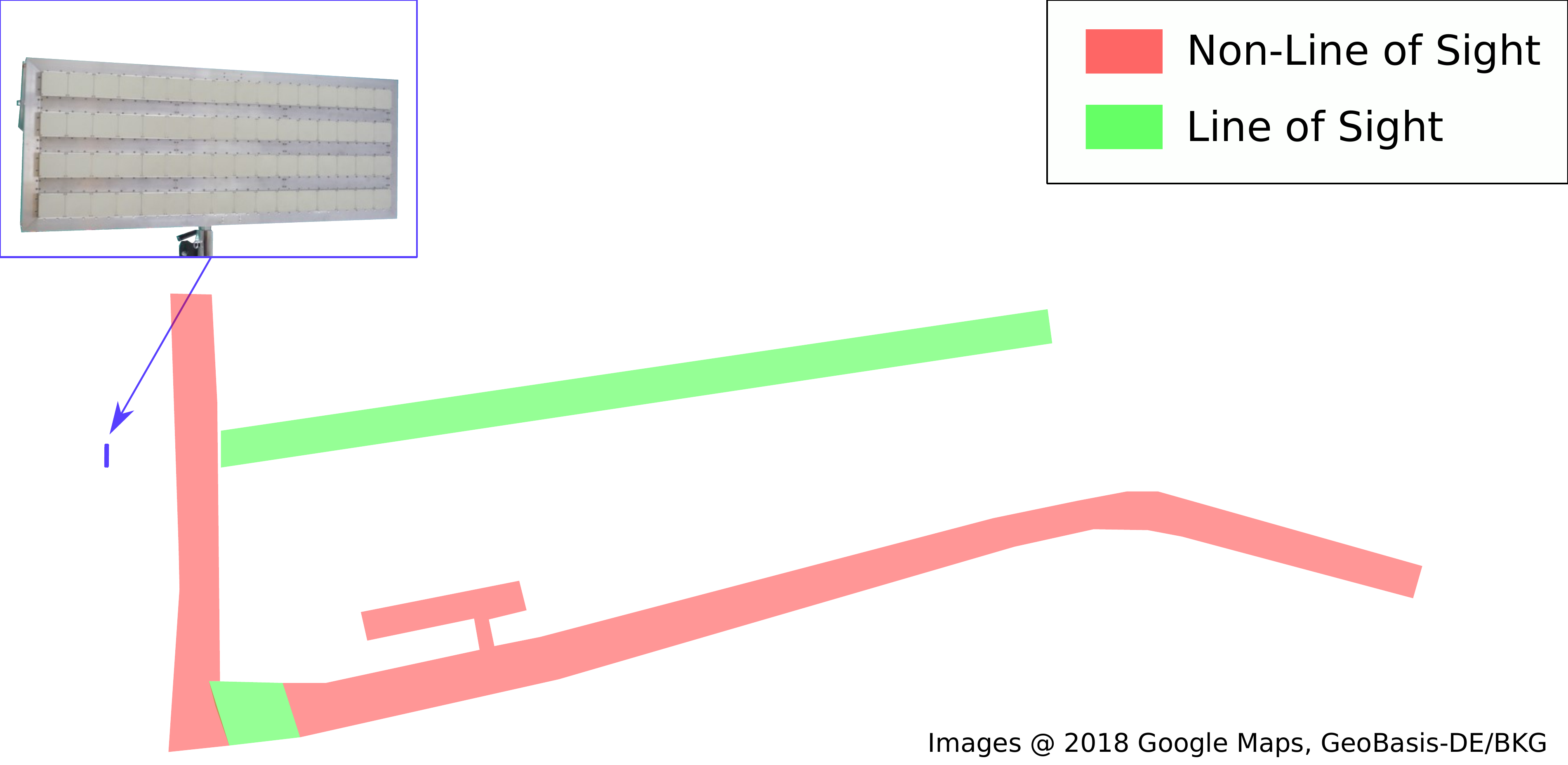}};
    \end{tikzpicture}
    \caption{Illustration showing the measurement site with the BS at a rooftop and LOS/NLOS conditions for the \ac{mt} locations~\cite{Hellings2019}.}
    \label{fig:meas_campaign}
\end{figure}

As described in~\cite{Hellings2019, Turan2022, Baur2023meas}, the measurement campaign was conducted at the Nokia campus in Stuttgart, Germany, in 2017. 
The scenario is displayed in Fig.~\ref{fig:meas_campaign}. 
It can be seen that the measurements were recorded in a street canyon surrounded by buildings producing a mix of \ac{los} and \ac{nlos} channels. 
The \ac{bs} is located on a rooftop approximately \SI{20}{m} above the ground with a \SI{10}{\degree} down-tilt.
It is a \ac{ura} with $\Nv=4$ vertical and $\Nh=16$ horizontal single polarized patch antennas and was adapted to match the \ac{3gpp} urban microcell propagation scenario---consequently, $N=64$.
The antenna spacing is $\lambda$ in the vertical and $\lambda/2$ in the horizontal direction, with $\lambda$ being the wavelength.

The single monopole receive-antenna representing the \ac{mt} was placed on a moving vehicle with a maximum speed of \SI{25}{km\per h}.
GPS was used to continuously establish synchronization between the \ac{bs} and \ac{mt}, yielding a channel realization every \SI{4}{\milli\meter} in space. 
The data was collected by a TSMW receiver and stored on a Rohde \& Schwarz IQR hard disk recorder.
The carrier frequency was \SI{2.18}{\giga\hertz}.
The \ac{bs} transmitted \SI{10}{\mega\hertz} \ac{ofdm} waveforms with $600$ subcarriers in \SI{15}{\kilo\hertz} spacing.
The pilots were sent continuously with a periodicity of \SI{0.5}{ms}, arranged in $50$ separate subbands of $12$ consecutive subcarriers. 
The channel was assumed to remain constant for one pilot burst. 
Channel realization vectors with $64$ coefficients per subband were extracted in a post-processing step. 
After the measurement campaign, the \ac{nmse} of the channel estimates was characterized to lie between $-20$ and \SI{-30}{dB}, reasonably motivating the assumption of possessing perfect \ac{csi} for the generative model training.


\subsection{Experiments}
\label{subsec:experiments}

This section presents the simulation results based on the real-world measurements from the previous section. 
We create a dataset of $\TR=400{,}000$ channel realizations for the training phases. 
For evaluation and testing purposes, we consider a $\TV=\TE=10{,}000$ channel realizations dataset.
The channels are normalized such that $\E[\|\vh\|^2]=N$. 
Furthermore, we define the \ac{snr} as $\frac{1}{\sigma^2}$ and calculate the \ac{nmse} as $\frac{1}{N\TE}\sum_{i=1}^\TE \| \vh_i - \hat{\vh}_i \|^2$ in the experiments, where $\vh_i$ and $\hat{\vh}_i$ mark the ground-truth channel and its estimate/reconstruction, respectively. 
We fit a \ac{gmm} with $K=64$ components for all experiments as considering more components did not notably improve the performance. 
For a detailed architecture and training description of the \ac{nn}-based methods, we refer to Appendix~\ref{app:implementation}. 
Overall, we chose comparable \ac{nn} architectures regarding layers and model weight numbers to ensure a fair comparison.

\subsubsection{Spectral Efficiency Analysis}
\label{subsubsec:se_analysis}

\pgfplotsset{every axis/.append style={font=\small}}

\begin{figure}[t]
    \centering
    \input{figures/cdf}
    \caption{Empirical \ac{se} CDFs for original samples from the \ac{rpe} or a generative model. For the corresponding W1D values, see Table~\ref{tab:eval_metric}.}
    \label{fig:cdf}
\end{figure}
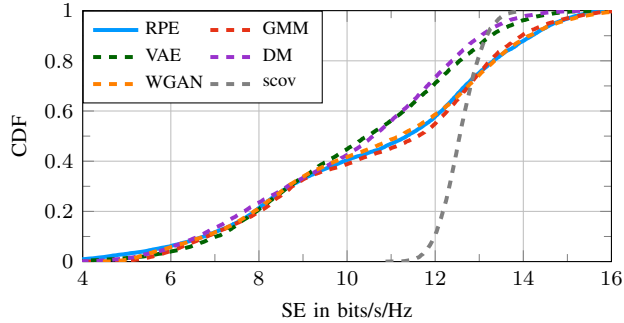

We begin with the \ac{se} analysis by comparing the empirical \ac{se} \acp{cdf} of samples from the \ac{rpe} and different generative models. 
To this end, we generate $10{,}000$ channel vector realizations with every generative model as described in Section~\ref{sec:prelim} and calculate the \ac{se} for every realization according to~\eqref{eq:se}. 
The result at an \ac{snr} of \SI{20}{dB} is displayed in Fig.~\ref{fig:cdf}.
To increase the validity of this evaluation method, we also show the result for the scov model, which is equivalent to a \ac{gmm} with one component. 
The figure highlights that the scov model cannot capture the \ac{se} characteristics of the \ac{rpe}.
In contrast, the \ac{gmm} and \ac{wgan} \ac{se} \acp{cdf} are very close to the \ac{rpe} \ac{cdf}. 
The \ac{vae} and \ac{dm} manage to reasonably approximate the \ac{rpe} \ac{cdf} for \acp{se} lower than \SI{10}{bits/s/Hz} and exhibit a moderate deviation for larger \acp{se}.

To quantify the discrepancies between the \ac{rpe} \ac{se} and the respective generative model \acp{se}, we also show the \ac{wd} in Table~\ref{tab:eval_metric} (second column). 
Every column entry represents the \ac{wd} between the \ac{rpe} \ac{se} and respective generative model \ac{se} distribution, e.g., the third row between the \ac{rpe} and \ac{vae}.     
Accordingly, the \ac{wgan} produces the best results in terms of the \ac{wd}, followed by the \ac{gmm}. 
Obviously, the scov has the worst \ac{wd} due to the bad \ac{se} reproduction. 
As explained in Appendix~\ref{app:implementation}, the \ac{wgan} is trained with entry-wise standardization, which is a reason for its good \ac{se} reproduction performance. 

\begin{table}[t]
    \centering
    \caption{Evaluation metrics for GM samples compared to RPE samples.}
    \label{tab:eval_metric}
    \begin{tabularx}{\tblw}{X|XXX}
        \hline
        Model & \multicolumn{1}{l}{W1D} & \multicolumn{1}{l}{TVD} & \multicolumn{1}{l}{MMD} \\ \hline
        GMM   & $5.3\cdot10^{-2}$       & $\mathbf{5.6\cdot10^{-2}}$       & $\mathbf{7.1\cdot10^{-5}}$       \\ \hline
        VAE   & $1.7\cdot10^{-1}$       & $8.3\cdot10^{-2}$       & $8.7\cdot10^{-3}$       \\ \hline
        DM    & $1.9\cdot10^{-1}$       & $8.5\cdot10^{-2}$       & $1.2\cdot10^{-2}$       \\ \hline
        WGAN   & $\mathbf{4.0\cdot10^{-2}}$       & $1.0\cdot10^{-1}$       & $4.1\cdot10^{-3}$       \\ \hline
        scov  & $7.0\cdot10^{-1}$       & $5.2\cdot10^{-1}$       & $3.6\cdot10^{-2}$       \\ \hline
    \end{tabularx}
\end{table}

\subsubsection{Codebook Fingerprinting}
\label{subsubsec:codebook_histograms}

\begin{figure*}[t]
     \begin{center}
        \hfill
        \subfigure[\leggmm\ and \legscov]{\input{figures/codebook_gmm}
        \label{fig:histogram_gmm}}
        \hfill
        \subfigure[\legvae]{\input{figures/codebook_vae}
        \label{fig:histogram_vae}}
        \hfill
     \end{center}
    \begin{center}
        \hfill
        \subfigure[\legdm]{\input{figures/codebook_dm}
        \label{fig:histogram_dm}}
        \hfill
        \subfigure[\leggan]{\input{figures/codebook_gan}
        \label{fig:histogram_gan}}
        \hfill
    \end{center}
    \caption{Codebook fingerprints for different GM samples compared to RPE samples. Each entry on the x-axis represents a codebook entry, and the corresponding relative frequency is shown on the y-axis. In (a), we show the relative frequencies for the complete codebook. The remaining histograms show only the first half for a more detailed exposition. For the corresponding TVD values, see Table~\ref{tab:eval_metric}.}
    \label{fig:codebook_histograms}
\end{figure*}
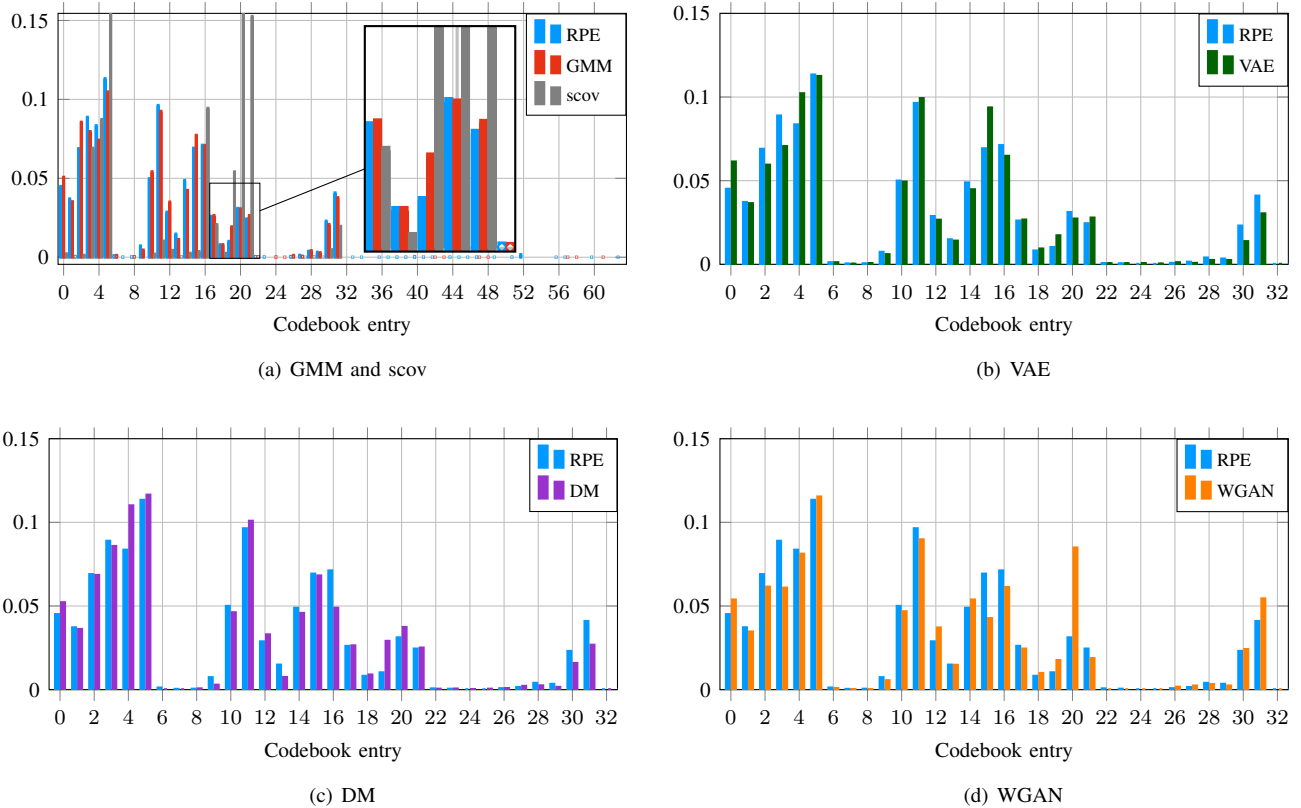

The \ac{se} analysis focuses on a generative model's channel norm reproduction capability. 
To also investigate the reproduction capability of the channel directions, we will make use of the proposed codebook fingerprinting evaluation routine. 
As for the \ac{se} analysis, we create $10{,}000$ channel realizations with every generative model for comparison with the \ac{rpe} test dataset. 
We utilize a codebook consisting of the Kronecker product of two \ac{dft} matrices motivated by the \ac{ura} at the \ac{bs}~\cite{Li2013}. 
More precisely, we adopt 
\begin{equation}
    \mathcal{C} = \left\{ \mf^{(\Nv)}_{C_1} \otimes \mf^{(\Nh)}_{C_2} \right\}
\end{equation}
as codebook where $\mf^{(\Nv)}_{C_1} \in \CC^{\Nv\times C_1}$ is a \ac{dft} matrix with $C_1$ columns, and $\mf^{(\Nh)}_{C_2} \in \CC^{\Nh\times C_2}$ analogously.
We further set $C_1=4$ and $C_2=16$ to yield a $B=6$ bits codebook with $C = C_1 C_2 = 64$ entries.

We plot the corresponding codebook fingerprints in Fig.~\ref{fig:codebook_histograms}, where each generative model is displayed in direct comparison to the \ac{rpe}.
In Fig.~\ref{fig:histogram_gmm}, we show the \ac{gmm} codebook fingerprint. 
As can be seen from the figure, the \ac{rpe} channel directions are almost wholly represented by the first codebook half, reflecting the inherent characteristics of the \ac{rpe}. 
The \ac{gmm} is very close to the \ac{rpe} codebook fingerprint. 
This impression is confirmed by the lowest \ac{tvd} in Table~\ref{tab:eval_metric}. 
Additionally, we present the scov codebook fingerprint in Fig.~\ref{fig:histogram_gmm}. 
The scov approach fails to replicate the \ac{rpe} codebook fingerprint and exhibits substantial deviations. 
Recalling that the \ac{tvd} lies between zero and one, the scov's \ac{tvd} value in Table~\ref{tab:eval_metric} is poor.

Moving on, in the remaining codebook fingerprints in Fig.~\ref{fig:codebook_histograms}, we only display the results of the first codebook half for a finer visualization. 
The \ac{vae} codebook fingerprint in Fig.~\ref{fig:histogram_vae} is again close to the \ac{rpe}, although there are some larger deviations from the \ac{rpe} after a detailed investigation. 
We observe a similar trend for the \ac{dm} in Fig.~\ref{fig:histogram_dm}, meaning the \ac{dm} codebook fingerprint also closely approaches the \ac{rpe}.
The \acp{tvd} of the \ac{vae} and \ac{dm} in Table~\ref{tab:eval_metric} validate their similar performance. 
At last, we investigate the \ac{wgan}'s codebook fingerprint in Fig.~\ref{fig:histogram_gan}. 
In this illustration, we observe some more significant deviations from the \ac{rpe}, e.g., at codebook entry $20$. 
The \ac{wgan}'s \ac{tvd} value in Table~\ref{tab:eval_metric} is the worst among the considered generative models. 
Therefore, the \ac{gmm} is best in replicating the directional information contained in the \ac{rpe}, followed by the \ac{vae}, \ac{dm}, and \ac{wgan} in that order.

\subsubsection{Application Cross-Check}
\label{subsubsec:eval_cross_check}

The third and last evaluation routine we investigate is the application cross-check from Section~{\ref{sec:eval_metrics}-\ref{subsec:cross-check}}. 
We begin with channel estimation as an application by considering the \ac{gmm}-based and \ac{vae}-based channel estimators as explained in Section~{\ref{sec:eval_metrics}-\ref{subsec:cross-check}}.
To this end, we first train all the considered generative models on the \ac{rpe} data as described in Appendix~\ref{app:implementation}. 
After the training, we generate a training, validation, and test dataset of the same size as for the \ac{rpe} with each generative model. 
The \ac{gmm}-based and \ac{vae}-based channel estimators are then trained on the generated data in the same manner as on the \ac{rpe} data, including usage of the same number of mixture components and architecture. 
The \ac{gmm} is fitted with $64$ components, and the \ac{vae} has the same architecture as in~\cite{Baur2023meas}. 
Ultimately, all channel estimators are evaluated on the same \ac{rpe} test dataset to compare their performance.

The performance comparison in terms of the \ac{nmse} is presented in Table~\ref{tab:cross_check} for an \ac{snr} of \SI{10}{dB}.
The \acp{nmse} in the table are scaled by $10^{-2}$.
Every row in the table is assigned to a channel estimator. 
Every column describes the data origin for the ``generated data-trained model'' in Fig.~\ref{fig:cross-check}. 
For instance, the entry in row ``VAE'' and column ``DM'' represents the \ac{nmse} of the \ac{vae}-based channel estimator that is trained with data from the \ac{dm} as a generative model. 
The \acp{nmse} in the column ``\ac{rpe}'' belong to the ``\ac{rpe} data-trained model'' in Fig.~\ref{fig:cross-check}, where the channel estimator in the corresponding table row is trained on the \ac{rpe} data. 
So, the other column entries should replicate the \acp{nmse} in the ``\ac{rpe}'' column for the best result.

\begin{table}[t]
\centering
\caption{NMSEs $(\cdot10^{-2})$ for different GM-based channel estimators.}
\label{tab:cross_check}
\begin{tabularx}{\linewidth}{l|XXXXXXX}
\hline
                                                & \multicolumn{7}{c}{Sample distribution}                                        \\ \hline
\parbox[t]{2mm}{\multirow{4}{*}{\rotatebox[origin=c]{90}{Estimator}}} & \multicolumn{1}{l|}{Method} & RPE & GMM           & scov & VAE  & DM   & WGAN  \\ \cline{2-8} 
                                                & \multicolumn{1}{l|}{GMM}    & $2.83$ & $\mathbf{2.90}$ & $5.64$ & $3.20$ & $3.40$ & $5.13$ \\ \cline{2-8} 
                                                & \multicolumn{1}{l|}{VAE}    & $3.38$ & $\mathbf{3.68}$ & $6.11$ & $3.97$ & $4.07$ & $4.85$ \\ \cline{2-8} 
                                                & \multicolumn{1}{l|}{LMMSE}   & $5.62$ & $\mathbf{5.62}$ & $5.65$ & $5.87$ & $6.01$ & $6.56$ \\ \hline
\end{tabularx}
\end{table}

Throughout all channel estimation experiments, the \ac{gmm} reproduces the ``\ac{rpe}'' \ac{nmse} best. 
For the ``GMM'' row, the strong performance can be partly attributed to fitting the \ac{gmm}-based estimator to \ac{gmm}-distributed data, which might be seen as an unfair advantage for the \ac{gmm} as a generative model. 
Thus, the result in the ``VAE'' column is important to highlight that the \ac{gmm} as a generative model also performs very well when the estimator is not \ac{gmm}-distributed, confirming the \ac{gmm}'s great performance also in this case. 
The second best generator with the \ac{gmm} and \ac{vae} as estimators is the \ac{vae} followed by the \ac{dm}.
In the last row of Table~\ref{tab:cross_check}, we also display the \ac{nmse} when using a simple \ac{lmmse} estimator with sample mean and covariance matrix as channel estimator, cf.~\eqref{eq:est_lmmse}. 
The \ac{lmmse} row highlights that the channel estimator should be powerful enough to learn a good estimate of the \ac{rpe} distribution since, in this row, using the scov method for generation performs very closely to the ``\ac{rpe}'' column, limiting the expressiveness of the result. 
Additionally, we present the performance of the \ac{gmm}-based channel estimator over the \ac{snr} for the considered generative models in Fig.~\ref{fig:gmm_cross_check}. 
We qualitatively observe the same behavior in this plot as for the ``GMM'' row in Table~\ref{tab:cross_check} since the \ac{gmm} performs again best, followed by the \ac{vae} and \ac{dm}. 
It is also visible that the \ac{wgan} shows the poorest performance among the generative models.

\begin{figure}[t]
    \centering
    \input{figures/gmm_cross_check}
    \caption{Application cross-check with the GMM channel estimator.}
    \label{fig:gmm_cross_check}
\end{figure}
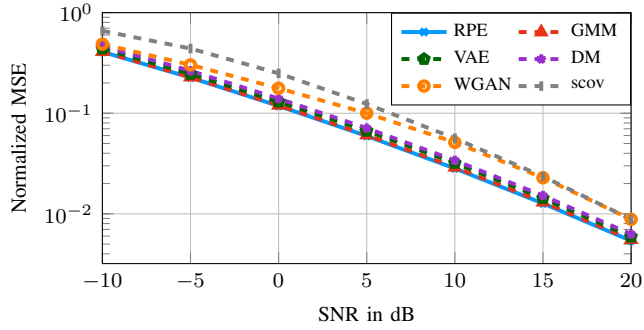

\begin{figure}[t]
    \centering
    \input{figures/ae_cross_check}
    \caption{Application cross-check with the AE for channel compression.}
    \label{fig:ae_cross_check}
\end{figure}
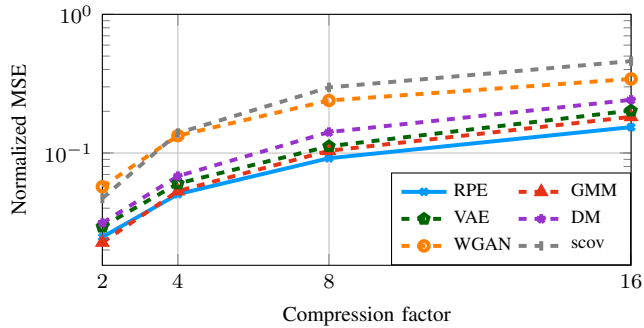

We continue with the cross-check with channel compression as an application. 
To this end, we exchange the ``application'' block in Fig.~\ref{fig:cross-check} with channel compression and leave the remainder as in the channel estimation check. 
We train an \ac{ae} where both the encoder and decoder are five-layer \acp{nn}. 
Every encoder layer consists of a convolutional layer with kernel size $9$ and $64$ convolutional channels, a batch normalization layer, and a ReLU activation function. 
The channel is downsampled until a compression factor of $\rho$ is reached while we consider $\rho\in\{2,4,8,16\}$. 
For the encoder input, the channel is first transformed to the beamspace, and then real and imaginary parts are the two input convolutional channels. 
The decoder is a symmetrical version of the encoder and upsamples the compressed channel until the original dimensionality is reached. 
This architecture is equivalent for every considered generative model training data. 
Fig.~\ref{fig:ae_cross_check} presents the simulation results for the channel compression cross-check. 
Again, the \ac{gmm} is closest to the \ac{rpe} performance, followed by the \ac{vae}, \ac{dm}, and \ac{wgan}. 
This result aligns with the results from the previous cross-check regarding channel estimation.

At last, we want to take a look at the rightmost column of Table~\ref{tab:eval_metric} where the \acp{mmd} between \ac{rpe} samples and the considered generative model samples are listed. 
It can be seen that the \ac{gmm} also achieves the best value in this comparison, with the \ac{wgan} and \ac{vae} being the second and third best, respectively. 
However, the table also highlights that the \ac{mmd} is intricate to interpret, and relying on the \ac{mmd} alone as an evaluation metric is insufficient. 
This can be deduced by comparing the \ac{mmd} value for the \ac{dm} and scov. 
The values lie closely together, although the \ac{dm} achieves significantly better generation results as the scov in all of our proposed evaluation routines. 
Moreover, the \ac{wgan} has the second best \ac{mmd} value, which might be misleading since the \ac{wgan} performed worst in the codebook fingerprinting and application cross-check among the considered generative models. 
These findings emphasize the importance of application-motivated evaluation routines for wireless data, as existing evaluation methods like the \ac{mmd} might produce misleading results. 

%% file: figures/cdf.tex
\begin{tikzpicture}
    \begin{axis}[
        width=\pltw,
        height = \plth,
        xlabel=SE in bits/s/Hz,
        ylabel=CDF,
        minor x tick num=1,
        minor y tick num=1,
        ymajorgrids=true,
        xmajorgrids=true,
        minor grid style={dashed},
        xmin=4, xmax=16,
        ymin=0, ymax=1,
        legend columns=2,
        legend cell align={left},
        legend style={font=\scriptsize},
        legend style={at={(0,1)},anchor=north west}]
        
        \addplot[RPE, each nth point=10]
        	table[x=cap_measure, y=cdf_measure, col sep=comma]
        	{data/cdf_gmm.txt};
        	\addlegendentry{\legrpe}
        \addplot[GMM, each nth point=10]
        	table[x=cap_gmm, y=cdf_gmm, col sep=comma]
        	{data/cdf_gmm.txt};
        	\addlegendentry{\leggmm}
        \addplot[VAE, each nth point=10]
        	table[x=cap_latent, y=cdf_latent, col sep=comma]
        	{data/cdf_vae.txt};
        	\addlegendentry{\legvae}
        \addplot[DM, each nth point=10]
        	table[x=cap_dm, y=cdf_dm, col sep=comma]
        	{data/cdf_dm.txt};
        	\addlegendentry{\legdm}
        \addplot[GAN, each nth point=10]
        	table[x=cap_gan, y=cdf_gan, col sep=comma]
        	{data/cdf_gan.txt};
        	\addlegendentry{\leggan}
        \addplot[scov, each nth point=10]
        	table[x=cap_scov, y=cdf_scov, col sep=comma]
        	{data/cdf_gmm.txt};
        	\addlegendentry{\legscov}
    
    \end{axis}
\end{tikzpicture}

%% file: figures/codebook_gmm.tex
\def\figuresize{2cm}
\pgfplotsset{scaled y ticks=false}
\begin{tikzpicture}[trim axis left, spy using outlines={rectangle, lens={scale=3}, height=3cm, width=2cm, connect spies}]
\begin{axis}[
    ybar=0.037cm,
    bar width=0.0001cm,
    height=\hsth,
    width=\hstw,
    xtick distance=4,
    ymin=0, ymax=0.15,
    yticklabels={-0.05,0,0.05,0.1,0.15},
    ymajorgrids=true,
    yminorgrids=true,
    xmajorgrids=true,
    xminorgrids=true,
    minor grid style={dashed},
    enlarge x limits=0.01,
    enlarge y limits=0.03,
    legend cell align={left},
    legend style={font=\scriptsize},
    xlabel={Codebook entry},
    legend style={at={(1,1)},anchor=north east}
    ]
    
    \addplot[RPEhist]
        table[y=measurement, x=entry, col sep=comma]{data/hist_gmm.txt};
        \addlegendentry{\legrpe}
    \addplot[GMMhist]
        table[y=gmm_full, x=entry, col sep=comma]{data/hist_gmm.txt};
        \addlegendentry{\leggmm}
    \addplot[scovhist]
        table[y=scov, x=entry, col sep=comma]{data/hist_gmm.txt};
        \addlegendentry{\legscov}

   \coordinate (c2) at (rel axis cs:0.5,0.5);
   \coordinate (c3) at (rel axis cs:0,0);

\end{axis}

\path (c3)--+(2.35,0.58) coordinate(C);  
\path (c2)--+(1.3,0) coordinate (B);  
\spy [black] on (C)
in node [] at (B);

\end{tikzpicture}

%% file: figures/codebook_vae.tex
\def\figuresize{2cm}
\begin{tikzpicture}[trim axis left, spy using outlines={rectangle, lens={scale=3}, height=3cm, width=2cm, connect spies}]
\begin{axis}[
    ybar=0.04cm,
    bar width=\histWidth,
    height=\hsth,
    width=\hstw,
    xtick distance=2,
    xmax=32,
    ymin=0, ymax=0.15,
    yticklabels={-0.05,0,0.05,0.1,0.15},
    ymajorgrids=true,
    yminorgrids=true,
    xmajorgrids=true,
    xminorgrids=true,
    minor grid style={dashed},
    enlarge x limits=0.02,
    legend cell align={left},
    legend style={font=\scriptsize},
    xlabel={Codebook entry},
    legend style={at={(1,1)},anchor=north east}
    ]
    
    \addplot[RPEhist]
        table[y=measurement, x=entry, col sep=comma]{data/hist_vae.txt};
        \addlegendentry{\legrpe}
    \addplot[VAEhist]
        table[y=latent, x=entry, col sep=comma]{data/hist_vae.txt};
        \addlegendentry{\legvae}

\end{axis}

\end{tikzpicture}

%% file: figures/codebook_dm.tex
\def\figuresize{2cm}
\begin{tikzpicture}[trim axis left, spy using outlines={rectangle, lens={scale=3}, height=3cm, width=2cm, connect spies}]
\begin{axis}[
    ybar=0.04cm,
    bar width=\histWidth,
    height=\hsth,
    width=\hstw,
    scaled ticks=false, 
    xtick distance=2,
    xmax=32,
    ymin=0, ymax=0.15,
    yticklabels={-0.05,0,0.05,0.1,0.15},
    ymajorgrids=true,
    yminorgrids=true,
    xmajorgrids=true,
    xminorgrids=true,
    minor grid style={dashed},
    enlarge x limits=0.02,
    legend cell align={left},
    legend style={font=\scriptsize},
    xlabel={Codebook entry},
    legend style={at={(1,1)},anchor=north east}
    ]
    
    \addplot[RPEhist]
        table[y=measurement, x=entry, col sep=comma]{data/hist_dm.txt};
        \addlegendentry{\legrpe}
    \addplot[DMhist]
        table[y=dm, x=entry, col sep=comma]{data/hist_dm.txt};
        \addlegendentry{\legdm}

\end{axis}

\end{tikzpicture}

%% file: figures/codebook_gan.tex
\def\figuresize{2cm}
\begin{tikzpicture}[trim axis left, spy using outlines={rectangle, lens={scale=3}, height=3cm, width=2cm, connect spies}]
\begin{axis}[
    ybar=0.04cm,
    bar width=\histWidth,
    height=\hsth,
    width=\hstw,
    xtick distance=2,
    xmax=32,
    ymin=0, ymax=0.15,
    yticklabels={-0.05,0,0.05,0.1,0.15},
    ymajorgrids=true,
    yminorgrids=true,
    xmajorgrids=true,
    xminorgrids=true,
    minor grid style={dashed},
    enlarge x limits=0.02,
    legend cell align={left},
    legend style={font=\scriptsize},
    xlabel={Codebook entry},
    legend style={at={(1,1)},anchor=north east}
    ]
    
    \addplot[RPEhist]
        table[y=measurement, x=entry, col sep=comma]{data/hist_gan.txt};
        \addlegendentry{\legrpe}
    \addplot[GANhist]
        table[y=gan, x=entry, col sep=comma]{data/hist_gan.txt};
        \addlegendentry{\leggan}

\end{axis}

\end{tikzpicture}

%% file: figures/gmm_cross_check.tex
\begin{tikzpicture}
    \begin{semilogyaxis}[
        width=\pltw,
        height = \plth,
        xlabel=SNR in dB,
        ylabel=Normalized MSE,
        ymajorgrids=true,
        xmajorgrids=true,
        minor grid style={dashed},
        xmin=-10, xmax=20,
        ymax=1,
        legend cell align={left},
        legend columns=2,
        legend style={font=\scriptsize},
        legend style={at={(1,1)},anchor=north east}]
        
        \addplot[RPEover]
        	table[x=snr, y=meas, col sep=comma]
        	{data/gmm_cross_check.txt};
        	\addlegendentry{\legrpe}
        \addplot[GMMover]
        	table[x=snr, y=gmm, col sep=comma]
        	{data/gmm_cross_check.txt};
        	\addlegendentry{\leggmm}
        \addplot[VAEover]
        	table[x=snr, y=vae, col sep=comma]
        	{data/gmm_cross_check.txt};
        	\addlegendentry{\legvae}
        \addplot[DMover]
        	table[x=snr, y=dm, col sep=comma]
        	{data/gmm_cross_check.txt};
        	\addlegendentry{\legdm}
        \addplot[GANover]
        	table[x=snr, y=gan, col sep=comma]
        	{data/gmm_cross_check.txt};
        	\addlegendentry{\leggan}
        \addplot[scovover]
        	table[x=snr, y=scov, col sep=comma]
        	{data/gmm_cross_check.txt};
        	\addlegendentry{\legscov}
    
    \end{semilogyaxis}
\end{tikzpicture}

%% file: figures/ae_cross_check.tex
\begin{tikzpicture}
    \begin{semilogyaxis}[
        width=\pltw,
        height = \plth,
        xlabel=Compression factor,
        ylabel=Normalized MSE,
        ymajorgrids=true,
        xmajorgrids=true,
        minor grid style={dashed},
        xmin=2, xmax=16,
        xtick={2,4,8,16},
        ymax=1,
        legend cell align={left},
        legend columns=2,
        legend style={font=\scriptsize},
        legend style={at={(1,0)},anchor=south east}]
        
        \addplot[RPEover]
        	table[x=snr, y=meas, col sep=comma]
        	{data/ae_cross_check.txt};
        	\addlegendentry{\legrpe}
        \addplot[GMMover]
        	table[x=snr, y=gmm, col sep=comma]
        	{data/ae_cross_check.txt};
        	\addlegendentry{\leggmm}
        \addplot[VAEover]
        	table[x=snr, y=vae, col sep=comma]
        	{data/ae_cross_check.txt};
        	\addlegendentry{\legvae}
        \addplot[DMover]
        	table[x=snr, y=dm, col sep=comma]
        	{data/ae_cross_check.txt};
        	\addlegendentry{\legdm}
        \addplot[GANover]
        	table[x=snr, y=gan, col sep=comma]
        	{data/ae_cross_check.txt};
        	\addlegendentry{\leggan}
        \addplot[scovover]
        	table[x=snr, y=scov, col sep=comma]
        	{data/ae_cross_check.txt};
        	\addlegendentry{\legscov}
    
    \end{semilogyaxis}
\end{tikzpicture}

%% file: content/conclusion.tex
\section{Concluding Remarks and Future Work}
\label{sec:conclusion}

In this work, we propose evaluation metrics and methods for generative models tailored toward wireless PHY layer data. 
The considered methods involve an \ac{se} comparison, a codebook fingerprinting, and an application cross-check. 
We consider popular generative models in the wireless literature in our evaluation based on real-world measurements, which include the \ac{gmm}, \ac{vae}, \ac{dm}, and \ac{wgan}. 
Our results indicate that solely relying on established metrics in the \ac{ml} literature, e.g., the \ac{mmd}, is insufficient and that wireless data requires evaluation routines motivated by wireless applications. 
To the best of our knowledge, our proposed evaluation metrics and methods provide the first framework for establishing evaluation routines for generative models in the wireless PHY layer.

We note that the results in this work depend on the system configuration and the considered \ac{rpe} represented by the measurement data. 
Depending on the system constraints as well as the dataset characteristics, the rankings of the generative models in Section~\ref{sec:results} might change. 
Indeed, the scope of this work is not to determine the best generative model but to establish general evaluation metrics and methods for generative models in the wireless PHY layer, which exhibit a connection to relevant applications. 
In future works, we want to extend the evaluation framework with the proposed routines by considering different typical PHY layer applications with various system constraints, such as a limited training dataset size, corrupted training data samples, model memory requirements, and online computational complexity.

%% file: content/app.tex
\section{Appendix}

\subsection{Implementation Details}
\label{app:implementation}

We implement all the \ac{nn} architectures with the help of \textit{PyTorch} and also follow the PyTorch nomenclature in describing the layers. 
That means the first argument in a convolutional layer refers to the input convolutional channels, the second to the output convolutional channels, etc. 
The same reasoning holds for all other layers.

\subsubsection{VAE Implementation}

The \ac{vae} architecture for generation is similar to the architecture in~\cite{Baur2023meas}. 
Thus, we parameterize $\mc_\vtheta(\vz)$ as a block-Toeplitz matrix in alignment with the \ac{ura}, i.e.,
\begin{equation}
    \mc_\vtheta(\vz) = \mq\herm \diag(\vc_\vtheta(\vz)) \mq, \quad \vc_\vtheta(\vz) \in \RR^{4N},
\end{equation}
where $\mq = \mf^{(2\Nv)}_{\Nv} \otimes \mf^{(2\Nh)}_{\Nh}$ is the Kronecker product of two two-times oversampled DFT-matrices.
Table~\ref{tab:arc-vae} separately displays the encoder and decoder architecture, where the leftmost column shows the layer number. 
The encoder receives $\vh$ as input, with the real and imaginary parts forming the two input convolutional channels. 
We additionally transform the encoder input to the beamspace as this benefitted the training in previous works~\cite{Baur2024}. 
The dimensionality of the latent space is set to $\NL=16$, and we use an exponential function to map to strictly positive values for $\vsig_\vphi$ and $\vc_\vtheta$. 
For the model weight optimization, we utilize the Adam optimizer with a learning rate of $0.0001$.
The model weights are optimized until saturation of the \ac{elbo} on the validation dataset is reached.

\begin{table}[t]
\centering
\caption{Architecture details of the implemented VAE.}
\label{tab:arc-vae}
\begin{tabular}{l|l|l}
\hline
\#                 & encoder                  & decoder                           \\ \hline
\multirow{2}{*}{0} & Conv1d(2, 16, 1, 1)      & Linear(16, 1225)                  \\ \cline{2-3} 
                   & -                        & Reshape(-1, 49, 25)               \\ \hline
\multirow{3}{*}{1} & Conv1d(16, 16, 11, 2, 1) & ConvTranspose1d(49, 28, 11, 2, 1) \\ \cline{2-3} 
                   & BatchNorm1d(16)          & BatchNorm1d(49)                   \\ \cline{2-3} 
                   & ReLU()                   & ReLU()                            \\ \hline
\multirow{3}{*}{2} & Conv1d(16, 28, 11, 2, 1) & ConvTranspose1d(28, 16, 11, 2, 1) \\ \cline{2-3} 
                   & BatchNorm1d(28)          & BatchNorm1d(16)                   \\ \cline{2-3} 
                   & ReLU()                   & ReLU()                            \\ \hline
\multirow{3}{*}{3} & Conv1d(28, 49, 11, 2, 1) & ConvTranspose1d(16, 3, 11, 2, 1)  \\ \cline{2-3} 
                   & BatchNorm1d(49)          & BatchNorm1d(3)                    \\ \cline{2-3} 
                   & ReLU()                   & ReLU()                            \\ \hline
\multirow{2}{*}{4} & Flatten(1)               & Flatten(1)                        \\ \cline{2-3} 
                   & Linear(1225, 32)         & Linear(747, 384)                  \\ \hline
\end{tabular}
\end{table}

\subsubsection{DM Implementation}

For the \ac{dm} implementation, we make use of the simulation code from~\cite{Fesl2024dm} and increase the number of model weights such that a similar number as in the \ac{vae} from Table~\ref{tab:arc-vae} is reached.
Table~\ref{tab:arc-dm} gives an overview of the \ac{dm} layers, which are split into a net-pre and a net-post. 
The convolutional layers perform a \textit{same} convolution. 
As displayed in~\cite[Fig. 1]{Fesl2024dm}, the network for the reverse process is split into two parts (net-pre and net-post here). 
After the net-pre, the positional embedding for $t$ is incorporated into the net-pre output. 
Then, the composition is processed by the net-post to yield $\vmu_\vtheta(\vh_t,t)$, cf.~\eqref{eq:dm_prob_reverse}. 
Moreover, the number of diffusion steps is set to $T=1000$ here.
Apart from that, the training procedure of the \ac{dm} is identical to~\cite{Fesl2024dm}, where the Adam optimizer with an initial learning rate of $0.0005$ is used that is step-wise decreased. 
The \ac{dm} is trained for up to $500$ epochs or until it saturates on the validation dataset.

\begin{table}[t]
\centering
\caption{Architecture details of the implemented DM.}
\label{tab:arc-dm}
\begin{tabularx}{\tblw}{l|X|X}
\hline
\#                 & net-pre             & net-post            \\ \hline
\multirow{2}{*}{0} & Conv1d(2, 26, 7)    & Conv1d(128, 102, 7) \\ \cline{2-3} 
                   & ReLU()              & ReLU()              \\ \hline
\multirow{2}{*}{1} & Conv1d(26, 51, 7)   & Conv1d(102, 77, 7)  \\ \cline{2-3} 
                   & ReLU()              & ReLU()              \\ \hline
\multirow{2}{*}{2} & Conv1d(51, 77, 7)   & Conv1d(77, 51 7)    \\ \cline{2-3} 
                   & ReLU()              & ReLU()              \\ \hline
\multirow{2}{*}{3} & Conv1d(77, 102, 7)  & Conv1d(51, 26, 7)   \\ \cline{2-3} 
                   & ReLU()              & ReLU()              \\ \hline
4                  & Conv1d(128, 102, 7) & Conv1d(26, 2, 7)    \\ \hline
\end{tabularx}
\end{table}

\subsubsection{WGAN Implementation}

The \ac{wgan} implementation follows the simulation code from~\cite{Doshi2022}, which includes considering the \ac{gp} for the training. 
We adapt the sizes of the layers to better fit the dimensionality of our channel realizations, which is $64$.
Table~\ref{tab:arc-gan} presents the corresponding architecture, where a latent dimension of $32$ is adopted, and \textit{same} convolutions are performed. 
Additionally, the channel vectors are transformed to the beamspace and standardized per entry as described in~\cite{Doshi2022}. 
The latter is separately done for the real and imaginary parts such that every entry has a mean zero and variance one since this procedure is reported to improve the \ac{wgan} performance~\cite{Srivastava2017}. 
We optimize the generator and discriminator weights with the RMSProp optimizer with a learning rate of $0.0001$ for up to $500$ epoch or until the validation loss saturates.

\begin{table}[t]
\centering
\caption{Architecture details for the implemented WGAN.}
\label{tab:arc-gan}
\begin{tabularx}{\tblw}{l|X|X}
\hline
\#                 & generator           & discriminator      \\ \hline
\multirow{4}{*}{0} & Linear(32, 2048)    & Conv1d(2, 16, 3)   \\ \cline{2-3} 
                   & ReLU()              & MaxPool1d(3, 2)    \\ \cline{2-3} 
                   & View(-1, 128, 16)   & LeakyReLU(0.2)     \\ \cline{2-3} 
                   & Upsample(None, 2.0) & Dropout(0.25)      \\ \hline
\multirow{4}{*}{1} & Conv1d(128, 128, 4) & Conv1d(16, 32, 3)  \\ \cline{2-3} 
                   & BatchNorm1d(128)    & MaxPool1d(3, 2)    \\ \cline{2-3} 
                   & ReLU()              & LeakyReLU(0.2)     \\ \cline{2-3} 
                   & Upsample(None, 2.0) & Dropout(0.25)      \\ \hline
\multirow{4}{*}{2} & Conv1d(128, 128, 4) & Conv1d(32, 64, 3)  \\ \cline{2-3} 
                   & BatchNorm1d(128)    & MaxPool1d(3, 2)    \\ \cline{2-3} 
                   & ReLU()              & LeakyReLU(0.2)     \\ \cline{2-3} 
                   & -                   & Dropout(0.25)      \\ \hline
\multirow{3}{*}{3} & Conv1d(128, 2, 4)   & Conv1d(64, 128, 3) \\ \cline{2-3} 
                   & -                   & LeakyReLU(0.2)     \\ \cline{2-3}
                   & -                   & Dropout(0.25)      \\ \hline
\multirow{2}{*}{4} & -                   & Flatten(1)         \\ \cline{2-3} 
                   & -                   & Linear(896, 1)     \\ \hline
\end{tabularx}
\end{table}